\documentclass[12pt,a4j]{article}
\usepackage[dvips]{graphicx}
\usepackage{enumerate}
\usepackage{amsmath}

\begin{document}
\baselineskip=5mm
\centerline{\large \bf Soliton and Periodic Solutions} \par
\centerline{\large \bf of the Short Pulse Model Equation}\par
\bigskip
\bigskip
\centerline{Yoshimasa Matsuno\footnote{{\it E-mail address}: matsuno@yamaguchi-u.ac.jp}}\par

\centerline{\it Division of Applied Mathematical Science, Graduate School of Science } \par
\centerline{\it and Engineering, Yamaguchi University, Ube, Yamaguchi 755-8611, Japan} \par

\bigskip
\bigskip
\centerline{\bf ABSTRACT} \par
The short pulse (SP) equation is a novel model equation describing the propagation of ultra-short optical
pulses in nonlinear media. This article reviews some recent results about the SP equation.
In particular, we focus our attention on its exact solutions.
By using a newly developed method of solution, we derive  multisoliton solutions as well as
1-and 2-phase periodic solutions and investigate their properties. \par
\bigskip
\leftline{\bf 1 INTRODUCTION}\par
In this article, we address  the following short pulse (SP) model
equation 
$$u_{xt}=u+{1\over 6 }(u^3)_{xx}, \eqno(1.1)$$
where $u=u(x,t)$ represents the magnitude of the electric field and subscripts $x$ and $t$
appended to $u$ denote  partial differentiation. 
The SP equation was proposed as a model nonlinear equation
describing the propagation of ultra-short optical pulses in nonlinear media [1]. It is
an alternative model equation to the cubic nonlinear Schr\"odinger (NLS) equation.
The basic assumption in deriving the NLS equation is a slowly varying amplitude approximation.
Hence,  as  discussed in
the context of self-focusing of ultra-short pulses in nonlinear media [2, 3], its validity 
would be violated if the pulse width becomes very short. A recent numerical analysis
reveals that as the pulse length shortens, the SP equation becomes a better approximation
to the solution of the Maxwell equation when compared with the prediction of the nonlinear
NLS equation [4]. Although the mathematical structure of the NLS equation has been studied extensively,
only a few results are known for the SP equation.  Here, we describe some recent results associated with the SP equation.
 In particular, we focus our attention on an exact method of solution, soliton and periodic
solutions and their properties.  \par
This article is organized as follows: In Sec. 2, we derive the SP equation starting with Maxwell equations
for the electric and magnetic fields.
In Sec. 3, an exact method of solution is developed for the SP equation which transforms the SP equation
to the integrable sine-Gordon (sG) equation through a hodograph transformation. In Sec. 4, the soliton solutions are
constructed which include the multiloop soliton and multibreather solutions.  Subsequently, the interaction process of solitons
is described in detail. In Sec. 5, the exact method is applied to obtaining 1- and 2-phase periodic solutions. Some properties of the
solutions are discussed as well as their long-wave limit. 
In Sec. 6, an alternative method of solution is introduced  which enables
us to construct a more general class of periodic solutions. Then, the 1- and 2-phase solutions are exemplified.
Section 7 is devoted to conclusion. \par
\bigskip
\leftline{\bf 2 SHORT PULSE EQUATION} \par
\leftline{\bf 2.1 Basic equations} \par

 The electric and magnetic fields ${\bf E}$ and ${\bf H}$ as well as the 
the electric and magnetic flux densities ${\bf D}$ and ${\bf B}$ are governed by the following set 
of equations
$$div\ {\bf D}=\rho,\quad div\ {\bf B}=0,\quad rot\ {\bf E}=-{\partial {\bf B}\over \partial t},
\quad rot\ {\bf H}={\bf j}+{\partial {\bf D}\over \partial t}, \eqno(2.1)$$
where $\rho$ and ${\bf j}$ are the electric charge and current densities, respectively.
We consider the one-dimensional propagation of the wave so that we can put
$${\bf E}=E_3(x,t){\bf e}_3,\quad {\bf H}=H_2(x,t){\bf e}_2, \eqno(2.2)$$
where ${\bf e}_2$ and ${\bf e}_3$ are unit vectors perpendicular to the $x$ axis.
We also assume the following relations
$${\bf D}=\epsilon_0 {\bf E}+{\bf P},\quad {\bf B}=\mu_0 {\bf H}, \eqno(2.3)$$
where ${\bf P}$ is the induced electric@polarization, $\epsilon_0$ is the vacuum permittivity and
$\mu$ is the vacuum permeability. In view of (2.2), Eqs. (2.1) and the first equation of (2.3)
are simplified to
$${\partial H_2\over \partial x}={\partial D_3\over \partial t},
\quad {\partial E_3\over \partial x}=\mu_0{\partial H_2\over \partial t},\eqno(2.4)$$
$$D_3=\epsilon_0 E_3+P_3, \eqno(2.5)$$
respectively. Combining (2.4) and (2.5), we obtain the equation for $E_3$. It reads in 
 form 
$$E_{xx}-{1\over c^2}E_{tt}=P_{tt}, \eqno(2.6)$$
where $E=E_3, P=\mu_0P_3$ and $c^2=(\epsilon_0\mu_0)^{-1}$.
The polarization $P$ can be split into the linear part $P_{\rm lin}$ and the nonlinear
part $P_{\rm nl}$ and it may be written in the form
$$P=P_{\rm lin}+P_{\rm nl}=\int^\infty_{-\infty}\chi^{(1)}(t-\tau)E(x,\tau)d\tau$$
$$+\int^\infty_{-\infty}\int^\infty_{-\infty}\int^\infty_{-\infty}\chi^{(3)}(t-\tau_1, t-\tau_2,  t-\tau_3)
E(x,\tau_1)E(x,\tau_2)E(x,\tau_3)d\tau_1d\tau_2d\tau_3, \eqno(2.7)$$
where $\chi^{(1)}$ and  $\chi^{(3)}$ are the susceptibilities. If we consider the propagation of
light with the wavelength between 1600nm and 3000nm, then the Fourier transform $\hat\chi^{(1)}$ of $\chi^{(1)}$ is found to
be well approximated by the relation $\hat\chi^{(1)}\simeq \hat\chi^{(1)}_0-\hat\chi^{(1)}_2 \lambda^2$ [1]. It follows from
this and the relation $\omega=2\pi c/\lambda$ that the linear equation for (2.6) written in Fourier transformed form becomes
$$\hat E_{xx}+{1+\hat\chi^{(1)}_0\over c^2}\omega^2\hat E-(2\pi)^2\hat\chi^{(1)}_2 \hat E=0. \eqno(2.8)$$
As for the nonlinear term in Eq. (2.6), we assume that the instantaneous contribution is dominant for the short and
small amplitude pulses. Under this situation, we can set $\chi^{(3)}(t-\tau_1, t-\tau_2,  t-\tau_3)
=\chi_3\delta(t-\tau_1)\delta(t-\tau_2)\delta(t-\tau_3)$ where $\chi_3$ is a constant. If we introduce this relation 
into the nonlinear term on the
right-hand side of (2.7), we obtain $P_{\rm nl}=\chi_3 E^3$, which combined with (2.8),   
 yields a single nonlinear wave equation for $E$
$$E_{xx}-{1\over c_1^2}E_{tt}={1\over c_2^2} E+\chi_3(E^3)_{tt}, \eqno(2.9)$$
where $c_1=c/\sqrt{1+\hat\chi^{(1)}_0}$ and $c_2=1/(2\pi\sqrt{\hat\chi^{(1)}_2})$. \par
\medskip
\leftline{\bf 2.2 Perturbation analysis} \par
Equation (2.9) describes the interactions between the left and right moving pulses. 
Since the pulses are very short, the interaction between them  would
give rise to a higher-order effect on the evolution of the waves. Consequently,  
we may address only the right moving pulses, for
instance. We use the multiple scale  method to derive the approximate equation by expanding $E$ as
$$E(x,t)=\epsilon E_0(\phi,X)+\epsilon^2 E_1(\phi, X) + \cdots, \eqno(2.10)$$
where $\epsilon$ is a small parameter which measures the shortness of the pulse relative to the
time scale determined by the resonance,  and $\phi$ and $X$ are the scaled variables defined by
$$\phi={t-{x\over c_1}\over \epsilon},\quad X=\epsilon x.\eqno(2.11)$$
If we introduce (2.10) with (2.11) into Eq. (2.9), we obtain, at the order $O(\epsilon)$, the following 
partial differential equation (PDE) for $E_0$:
$$-{2\over c_1}{\partial^2E_0\over \partial\phi\partial X}={1\over c_2^2} E_0+\chi_3{\partial^2E_0^3
\over\partial\phi^2}. \eqno(2.12)$$
After an appropriate change of the variables,  we arrive at the normalized form of the SP equation (1.1). \par
\medskip
\leftline{\bf 2.3 Remarks}\par
\noindent 1. The SP equation has been derived for the first
time in an attempt to construct integrable differential equations associated with pseudospherical surfaces [5].
Sch\"afer and Wayne  rederived it starting from  Maxwell's equations of electric field
in the fiber as described in this section. See also a prior work due to Alterman and Rauch who
discuss the breakdown of the slowly varying envelope approximation and perform an asymptotic analysis
for a new type of nonlinear evolution equation [6]. \par
\noindent 2. The integrability of the SP equation has been established from various mathematical points 
of view [5, 7-10]. \par
\noindent 3. There exist several analogous equations to the SP equation which have been proven to be completely
integrable. We write one of them in the form
$$u_{xt}=\alpha u+{1\over 2}(1-\beta)u_x^2-uu_{xx}. \eqno(2.13)$$ 
When $\beta=2$, Eq. (2.13) becomes the short-wave model for the Camassa-Holm equation  while
when $\beta=4$, it reduces to the the short-wave model for the Degasperis-Procesi equation  and
the Vakhnenko equation.  The general multisoliton solutions of these equations have been
obtained in parametric forms [11].  \par
\bigskip
\leftline{\bf 3 METHOD OF EXACT SOLUTION} \par
\leftline{\bf 3.1 Reduction to the sine-Gordon equation}\par
Here we develop an analytical method for solving the SP equation which
employs the hodograph transformation to reduce  it to the completely
integrable sG equation [12].
We first introduce  the new dependent variable $r$ 
$$r^2=1+u_x^2, \eqno(3.1)$$
to transform the SP equation (1.1) into the form of conservation law
$$ r_t=\left({1\over 2}u^2r\right)_x. \eqno(3.2)$$
We then define the hodograph transformation $(x ,t) \rightarrow (y, \tau)$ by means of
$$dy=rdx+{1\over 2}u^2rdt, \quad d\tau=dt, \eqno(3.3a)$$
or equivalently 
$${\partial\over \partial x}=r {\partial\over \partial y}, \ {\partial\over \partial t}=
{\partial\over \partial \tau}+{1\over 2}u^2r {\partial\over \partial y}. \eqno(3.3b)$$
In terms of the new variables $y$ and $\tau$, (3.1) and  (3.2) are recast into
$$r^2=1+r^2u_y^2,\eqno(3.4)$$

$$r_\tau=r^2uu_y, \eqno(3.5)$$
respectively.  Furthermore, we define the  variable $\phi$ by
$$u_y=\sin \phi, \quad \phi=\phi(y,\tau).\eqno(3.6)$$
Inserting (3.6) into (3.4) gives
$${1\over r}=\cos \phi. \eqno(3.7)$$
It follows from (3.5)-(3.7) that

$$u=\phi_\tau. \eqno(3.8)$$
Finally, if we substitute (3.8) into (3.6), we find that $\phi$ obeys the following sG
equation
$$\phi_{y\tau}=\sin \phi. \eqno(3.9)$$
This form of the sG equation will be used to construct soliton solutions of the SP equation.
For the periodic solutions, it is appropriate  to introduce the two independent
 phase variables $\xi$ and $\eta$ according to
 $$\xi=ay+{\tau\over a}+\xi_0, \eqno(3.10a)$$
 $$\eta=ay-{\tau\over a}+\eta_0, \eqno(3.10b)$$
 where $a(\not=0), \xi_0$ and $\eta_0$ are arbitrary constants. 
In terms of the new variables, the sG equation (3.9) is transformed to 
 $$\phi_{\xi\xi}-\phi_{\eta\eta}=\sin \phi, \ \phi=\phi(\xi, \eta). \eqno(3.11)$$
 \leftline{\bf 3.2 Parametric representation of the solution} \par
 \medskip
 The solution $u$ has a parametric representation given by (3.8). To be more specific
 $$u(y,\tau)=\phi_\tau. \eqno(3.12)$$
 To obtain the parametric representation of the coordinate $x$, we
note from (3.3b) that the inverse mapping $(y,\tau) \rightarrow (x,t)$ is   governed 
by the system of linear  PDE for $x=x(y, \tau)$
$$x_y={1\over r}, \quad x_\tau=-{1\over 2}u^2. \eqno(3.13)$$ 
Since the integrability of the above system of equations is assured automatically by Eq. (3.5),
 we are able to integrate (3.13) immediately to obtain
$$x(y,\tau)=\int\cos \phi\ dy+c, \eqno(3.14)$$
where $c$ is an integration constant. \par
\medskip
\leftline{\bf 3.3 Criterion for the single-valued solutions} \par
As will be demonstrated later, most of the parametric solutions (3.12) and (3.14)  become  multivalued functions 
for both the soliton and periodic solutions. The single-valued functions are
 particularly useful in application to the real physical problem such as the
 propagation of nonlinear short pulses in an optical fiber.
 A criterion for single-valued functions may be obtained
 simply by requiring that $u_x$ exhibits no singularities.  It follows from (3.3b), (3.6)
and (3.7) that $u_x=\tan \phi$. Thus, if
$$ -{\pi\over 2}< \phi < {\pi\over 2}, \pmod{\pi}, \ (-\sqrt{2}+1<\tan{\phi\over 4}<\sqrt{2}-1). \eqno(3.15)$$
then the parametric solutions (3.12) and (3.14) would become single-valued functions for
all values of $x$ and $t$. \par
\medskip
\leftline{\bf 3.4 Remark} \par
The reduction of the SP equation to the sG equation has also been established through the chain of transformations [8]. \par
\bigskip
\leftline{\bf 4 SOLITON SOLUTIONS}\par
\leftline{\bf 4.1 Parametric representation of the $N$-soliton solution}\par
In the context of the sG model, the soliton solutions are called kinks or breathers. These
solutions are reduced from the soliton solutions by specifying the parameters such as the amplitude and
the phase. Here we present the parametric representation for the $N$-soliton solution of the SP
equation [12].\par
 The general $N$-soliton solution of the sG equation can be written in a compact form as [13]
 $$\phi=2{\rm i}\ln {f^\prime\over f},\eqno(4.1)$$
 with
$$f=\sum_{\mu=0,1}{\rm exp}\left[\sum_{j=1}^N\mu_j\left(\xi_j+{\pi\over 2}{\rm i}\right)
+\sum_{1\le j<k\le N}\mu_j\mu_k\gamma_{jk}\right], \eqno(4.2a)$$
  $$f^\prime=\sum_{\mu=0,1}{\rm exp}\left[\sum_{j=1}^N\mu_j\left(\xi_j-{\pi\over 2}{\rm i}\right)
+\sum_{1\le j<k\le N}\mu_j\mu_k\gamma_{jk}\right], \eqno(4.2b)$$
$$\xi_j=p_jy+{1\over p_j}t+\xi_{j0}, \quad (j=1, 2, ..., N),\eqno(4.2c)$$
$${\rm e}^{\gamma_{jk}}=\left({p_j-p_k\over p_j+p_k}\right)^2, \quad (j, k=1, 2, ..., N; j\not=k).
\eqno(4.2d)$$
Here, $p_j$ and $\xi_{j0}$ are arbitrary complex parameters satisfying the conditions $p_j\not=\pm p_k$
for $j\not= k$, ${\rm i}=\sqrt{-1}$ and $N$ is an arbitrary positive integer. The notation $\sum_{\mu=0,1}$
implies the summation over all possible combination of $\mu_1=0, 1, \mu_2=0, 1, ..., 
\mu_N=0, 1$. Note in (4.2c) that  the variable $\tau$ is replaced by $t$  taking into account (3.3a).
This convention will be  used  in the following.  The $\tau$-functions $f$ and $f^\prime$ play an essential
role in constructing soliton solutions, as will be seen below. They satisfy the following system of
bilinear equations 
$$ff_{yt}-f_yf_t={1\over 4}(f^2-{f^\prime}^2),\eqno(4.3a)$$
$$f^\prime f^\prime_{yt}-f^\prime_yf^\prime_t={1\over 4}({f^\prime}^2-f^2).\eqno(4.3b)$$
We obtain from (4.1) and (4.3) the important relation 
$$\cos \phi=1-2(\ln f^\prime f)_{yt}.\eqno(4.4)$$
 Introducing (4.4) into (3.14) and integrating with respect to $y$ yield the parametric form of the
 coordinate $x$
$$x(y,t)=y-2(\ln f^\prime f)_t+c, \eqno(4.5)$$
where $c$ is an integration constant depending generally on $t$.
It also follows from (3.12) and  (4.1) that
$$u(y,t)=2{\rm i}\left(\ln {f^\prime\over f}\right)_t,\eqno(4.6)$$
which, combined with (4.6), gives the parametric representation of the $N$-soliton solution of the
SP equation. 
To complete the solution, one must determine the time dependence of $c$. To this end,
 we substitute (4.5) and (4.6) into the
second equation of (3.13) and obtain the bilinear equation for $f$ and $f^\prime$
$$ff^\prime_{tt}-2f^\prime_tf_t+f^\prime f_{tt}={1\over 2}c^\prime(t)f^\prime f. \eqno(4.7)$$
One can show that the $\tau$-functions $f$ and $f^\prime$ from (4.2) vanish the left-hand side of
(4.7). Consequently, $c^\prime(t)=0$. Thus, in the case of soliton solutions, the constant $c$
does not depend on $t$. 
 In the following discussion, we consider real $u$ so that we take $f^\prime=f^*$ (complex
conjugate of $f$). As will be seen later, this restriction imposes certain conditions on the parameters
$p_j$ and $\xi_{j0}\ (j=1, 2, ..., N)$.  The loop soliton and breather solutions of the SP
equation are special classes of the general $N$-soliton solutions given by (4.5) and (4.6). \par
If we use (4.1) with $f^\prime=f^*$ and the formula ${\rm i\ ln}(f^*/f)=2\,{\rm tan}^{-1}({\rm Im}\ f/{\rm Re}\ f)$,
the criterion (3.15) for the single-valued solutions can be rewritten as
$$-\sqrt{2}+1<{{\rm Im}\ f\over {\rm Re}\ f}<\sqrt{2}-1. \eqno(4.8)$$
\par
 \leftline{\bf 4.2 Loop soliton solutions}\par
 \leftline{\bf 4.2.1 1-loop soliton solution} \par
Various solutions can be obtained for the SP equation by specifying the parameters $p_j$ 
and $\xi_{j0} (j=1, 2, ..., N)$ in (4.2). 
The loop (antiloop) soliton
 solutions arise from the kink (antikink) solutions of the sG equation.  
 Let $m$ and $N-m$ be the number of positive and negative $p_j$, respectively.
Then, the corresponding soliton solution would describe the interaction of the $m$ loop solitons
and $N-m$ antiloop solitons. Here we address the simplest  1-loop soliton solution which is of fundamental
 importance in discussing the properties of the general $N$-loop  soliton solution.
In this case, (4.2) is written as
$$f=1+{\rm i}{\rm e}^{\xi_1},\quad \xi_1=p_1y+{1\over p_1}t+\xi_{10}, \eqno(4.9)$$
\begin{center}
\includegraphics[width=10cm]{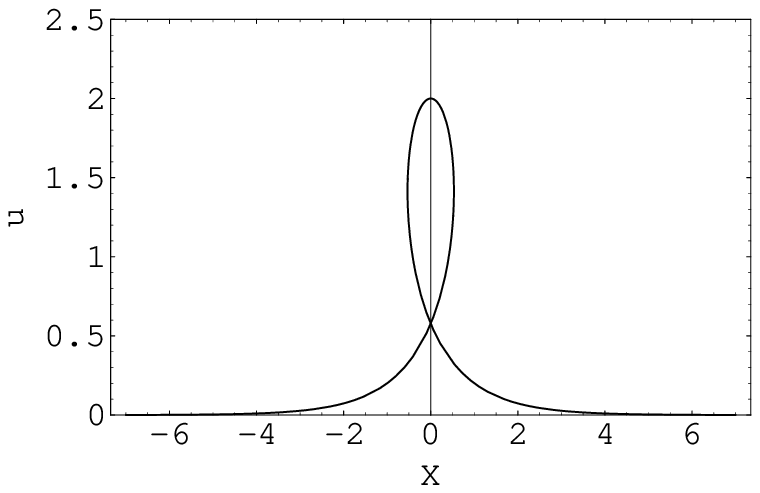}
\end{center}
\centerline{{\bf Fig. 1}: A typical profile of the 1-loop soliton solution.} \par
\bigskip
\noindent and $f^\prime=f^*$. Substituting (4.9) into (4.5) and (4.6), we obtain
$$u(y,t)={2\over p_1}{\rm sech} \xi_1, \eqno(4.10a)$$
$$x(y,t)=y-{2\over p_1}\tanh \xi_1 +d_1, \eqno(4.10b)$$
where $d_1=c-(2/p_1)$, $p_1>0$ and $\xi_{10}$ is a real constant. 
If we introduce a new variable $X\equiv x+c_1t -x_{10}$ with $c_1=1/p_1^2$ and $x_{10}=-\xi_{10}/p_1$,
then we can parameterize $x(y, t)$ by a single variable $\xi_1$. To be more specific, it reads
$$X={\xi_1\over p_1}-{2\over p_1}\tanh \xi_1 +d_1. \eqno(4.11)$$
It follows from (4.10a) and (4.11) that
$${du\over dX}={\sinh\ \xi_1\over 2-\cosh^2\xi_1}. \eqno(4.12)$$
We see from (4.12) that $du/dX$ changes sign three times and goes infinity at
$\xi_1=\pm \cosh^{-1}2$. Thus, the parametric solutoin exhibits singularities which has a form of single
loop. The multi-valued feature of the solution is also confirmed by applying the criterion (4.8). 
In fact, it follows from (4.9) that ${\rm Re}\ f =1$ and ${\rm Im}\ f={\rm e}^{\xi_1}$ and hence  (4.8)
cannot be satisfied for arbitrary values of $y$ and $t$.  
Figure 1 shows a typical profile of the 1-loop soliton solution with
the parameters $p_1=1, d_1=0$. 
The loop soliton propagates to the left (i.e., negative $x$ direction) at
a constant velocity $c_1$.  If we define the amplitude $A_1$ of the loop soliton
by $2/p_1$ (maximum value of $u$), then $c_1=A_1^2/4$. Thus, the large loop
soliton  moves more rapidly than the small loop soliton, indicating the typical
solitonic behavior. 

\par
\medskip
 \leftline{\bf 4.2.2 $N$-loop soliton solution}\par
 The general $N$-loop soliton solution of the SP equation arises from (4.5) and (4.6)
 by taking the parameters $p_j (j=1, 2, ..., N)$ positive and $\xi_{j0} (j=1, 2, ..., N)$ real. 
 We first investigate the asymptotic
 behavior of the solution for large time and show that it is represented by a
 superposition of $N$-loop solitons. The procedure for deriving the
 large time asymptptics can be performed straightforwardly by investigating the behavior of the
 $\tau$-functions given by (4.2). Hence, we omit the detail and describe only the result.
 To this end, we put
$c_j=1/p_j^2$ and order the magnitude of the velocity of each loop soliton
as $c_1>c_2> ...>c_N$.  We observe the interaction of $N$ loop solitons in a moving frame
with a constant velocity $c_n$. We  take the limit $t \rightarrow -\infty$
with the phase variable $\xi_n$ being fixed. We then find the following asymptotic form of $u$ and
$x$:
$$u \sim {2\over p_n}{\rm sech}\left(\xi_n+\delta_n^{(-)}\right), \eqno(4.13a)$$
$$x \sim y-{2\over p_n}\tanh\left(\xi_n+\delta_n^{(-)}\right)-4\sum_{j=n+1}^N{1\over p_j}
-{2\over p_n}+c. \eqno(4.13b)$$
where
$$\delta_n^{(-)}=\sum_{j=n+1}^N\ln\left({p_n-p_j\over p_n+p_j}\right)^2. \eqno(4.13c)$$
The corresponding asymptotic forms for $t \rightarrow +\infty$ are given by
$$u \sim {2\over p_n}{\rm sech}\left(\xi_n+\delta_n^{(+)}\right), \eqno(4.14a)$$
$$x \sim y-{2\over p_n}\tanh\left(\xi_n+\delta_n^{(+)}\right)-4\sum_{j=1}^{n-1}{1\over p_j}
-{2\over p_n}+c, \eqno(4.14b)$$
with
$$\delta_n^{(+)}=\sum_{j=1}^{n-1}\ln\left({p_n-p_j\over p_n+p_j}\right)^2. \eqno(4.14c)$$
\par
Let $x_c$ be the center position of the $n$th loop soliton in the $(x,t)$ coordinate
system. It then follows from (4.13) and (4.14) that
$$x_c+c_nt-x_{n0} \sim -{\delta_n^{(-)}\over p_n}-4\sum_{j=n+1}^N{1\over p_j}+d_n,\ (t \rightarrow -\infty)\eqno(4.15a)$$
$$x_c+c_nt-x_{n0} \sim -{\delta_n^{(+)}\over p_n}-4\sum_{j=1}^{n-1}{1\over p_j}+d_n,\ (t \rightarrow +\infty),\eqno(4.15b)$$
where $x_{n0}=-\xi_{n0}/p_n$ and $d_n=c-2/p_n$ are phase constants.
In view of the fact that all the loop solitons propagate to the left, we can define the
phase shift of the $n$th loop soliton as
$$\Delta_n=x_c(t\rightarrow -\infty)-x_c(t\rightarrow +\infty). \eqno(4.16)$$
This quantity is evaluated using (4.13c), (4.14c) and (4.15) to give
$$\Delta_n={1\over p_n}\left\{\sum_{j=1}^{n-1}\ln\left({p_n-p_j\over p_n+p_j}\right)^2
-\sum_{j=n+1}^N\ln\left({p_n-p_j\over p_n+p_j}\right)^2\right\}$$
$$+4\left(\sum_{j=1}^{n-1}{1\over p_j}-\sum_{j=n+1}^N{1\over p_j}\right),
\quad (n = 1, 2, ..., N). \eqno(4.17)$$
Note that the first term on the right-hand side of (4.17) coincides with the formula
for the phase shift arising from the interaction of $N$ kinks of the
sG equation. On the other hand, the second term arises due to the coordinate
transformation (3.3). The latter changes the characteristics of the
interaction process of  loop solitons  substantially when compared with those of the sG
kinks. \par
\medskip
\noindent {\bf 4.2.3 2-loop soliton solution}\par
The $\tau$-functions $f$ and $f^\prime$ for the 2-loop soliton solution are written as
$$f=1+{\rm i}{\rm e}^{\xi_1}+{\rm i}{\rm e}^{\xi_2}-\gamma {\rm e}^{\xi_1+\xi_2},\eqno(4.18a)$$
and $f^\prime=f^*$ with
$$\gamma=\left({p_1-p_2\over p_1+p_2}\right)^2. \eqno(4.18b)$$
\begin{center}
\includegraphics[width=10cm]{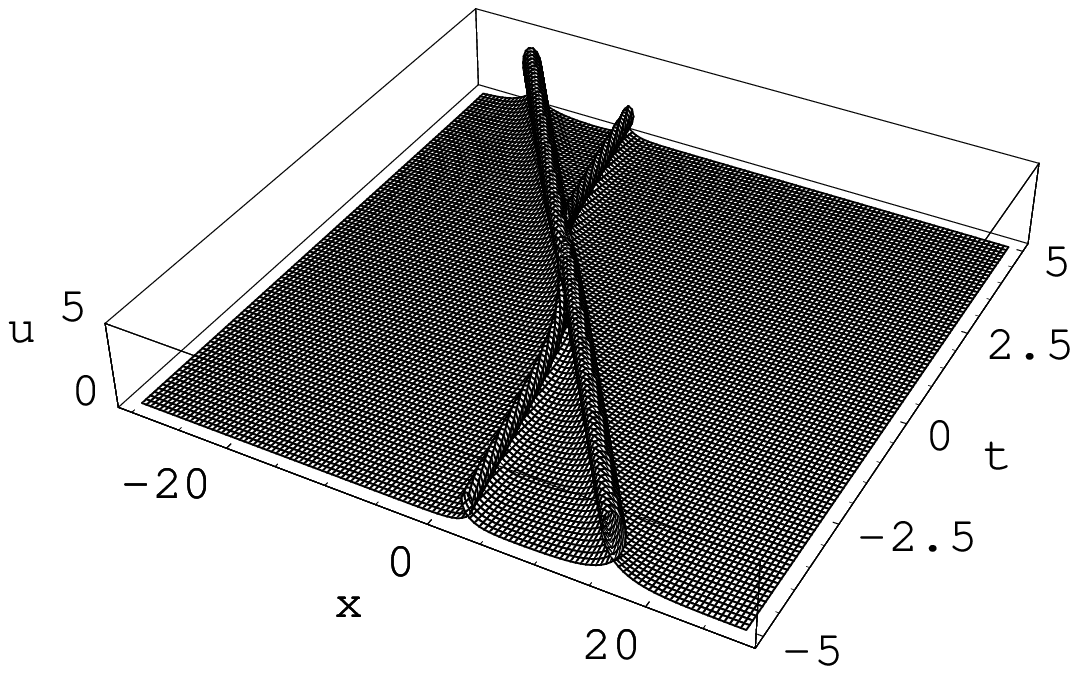}
\end{center}
\centerline{{\bf Fig. 2}: The interaction of two loop solitons.} \par
\bigskip 
\begin{center}
\includegraphics[width=10cm]{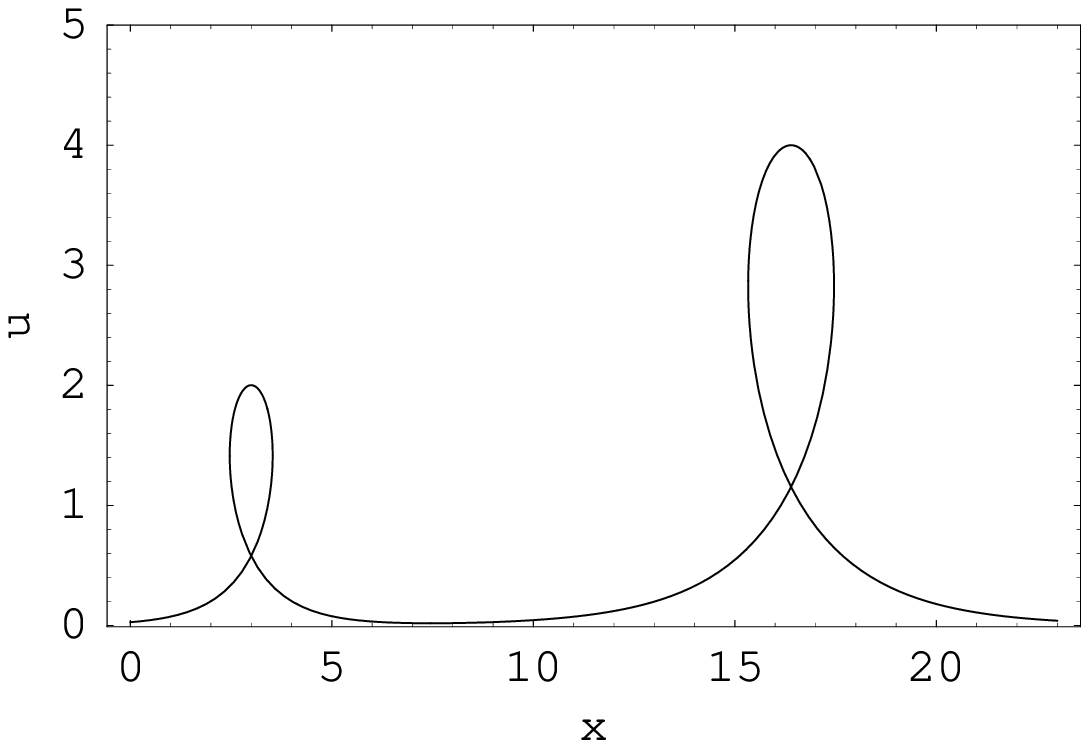}
\end{center}
\centerline{{\bf Fig. 3}: The profile of the 2-loop soliton solution.} \par
\bigskip
The parametric representation of the solution is then given by (4.5), (4.6) and (4.18). It reads
$$u(y,t)={2\sqrt{\gamma}\over p_1p_2}{(p_1+p_2)\cosh \psi_1 \cosh \psi_2+(p_1-p_2)
\sinh \psi_1 \sinh \psi_2\over \cosh^2\psi_1+\gamma \sinh^2\psi_2}, \eqno(4.19a)$$
$$x(y,t)=y+{1\over p_1p_2}{(p_1-p_2)\sinh 2\psi_1 -\gamma(p_1+p_2)\sinh 2\psi_2 \over 
\cosh^2\psi_1+\gamma \sinh^2\psi_2}-{2(p_1+p_2)\over p_1p_2}+c, \eqno(4.19b)$$
where we have put
$$\psi_1={1\over 2}(\xi_1-\xi_2), \quad \psi_2={1\over 2}(\xi_1+\xi_2)+{1\over 2}\ln \gamma, \eqno(4.19c)$$
for simplicity.
The positive parameters $p_1$ and $p_2$ are assumed to satisfy the condition $p_2>p_1$. 
Figure 2 shows the interaction of two loop solitons with the parameters given by $p_1=0.5, p_2=1.0,
c=\xi_{10}=\xi_{20}=0$. \par
Figure 3 shows the profile of the 2-loop soliton solution at $t=-5$ with the same parameters as those of Fig. 2.
For the 2-loop soliton case, formulas (4.17) for the phase shift are written as
$$\Delta_1=-{1\over p_1}\ln\left({p_1-p_2\over p_1+p_2}\right)^2-{4\over p_2},\eqno(4.20a)$$
$$\Delta_2={1\over p_2}\ln\left({p_1-p_2\over p_1+p_2}\right)^2+{4\over p_1}.\eqno(4.20b)$$
Figure 4 plots $p_1\Delta_1$ and $p_1\Delta_2$ as a function of $s(\equiv p_1/p_2)$.
 Thus, the large loop soliton always exhibits a positive phase shift whereas
the small one exhibits a positive phase shift for $0<s<s_c $ and a negative
phase shift for $s_c<s<1$ where $s_c$ is a solution of the    transcendental 
equation $\Delta_2=0$ and is given by $s_c=0.834$.  \par
\bigskip 
\begin{center}
\includegraphics[width=10cm]{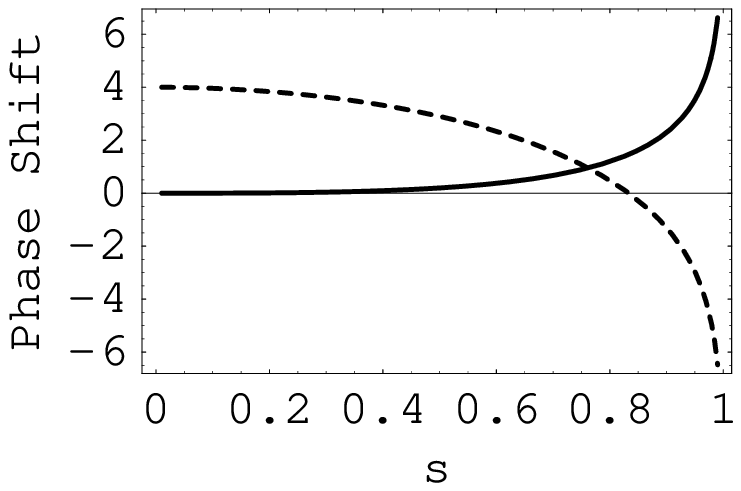}
\end{center}
\noindent{\bf Fig. 4}: The phase shift $p_1\Delta_1$ and $p_1\Delta_2$ as a function of $s$.  The
solid (broken) line represents the phase shift of the large (small) loop soliton.  \par
\bigskip
\noindent {\bf 4.2.4 Loop-antiloop soliton solution}\par
The solution representing the interaction of a loop soliton and an antiloop soliton arises if we choose
$p_1>0$ and $p_2<0$ with $p_1<|p_2|$ in the 2-soliton $\tau$-function (4.2).  
The parametric solution takes exactly the same form as that of the 2-loop soliton solution (4.19).\par
\bigskip
\begin{center}
\includegraphics[width=10cm]{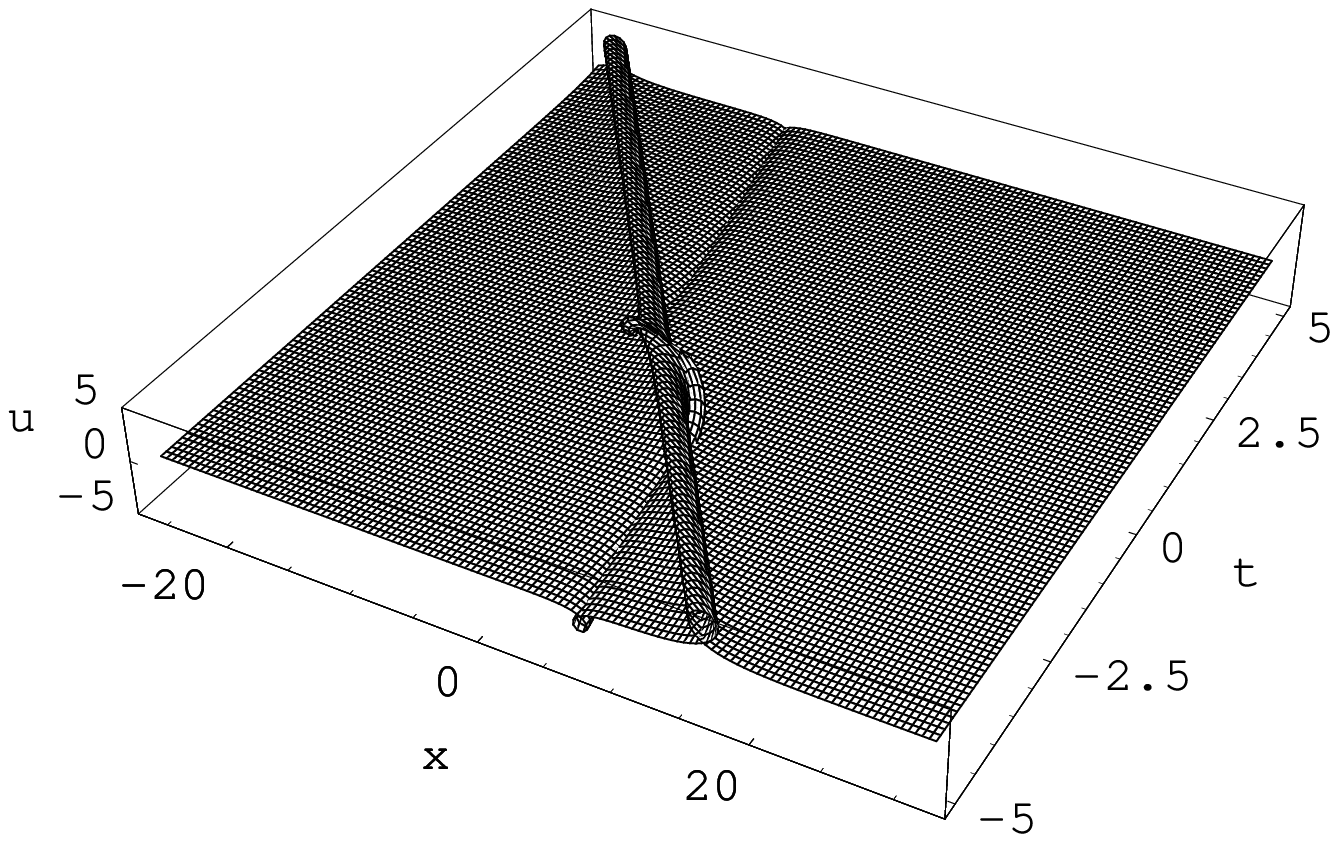}
\end{center}
\centerline{{\bf Fig. 5}: The interaction of a loop soliton and an antiloop soliton.} \par
\bigskip 
\begin{center}
\includegraphics[width=10cm]{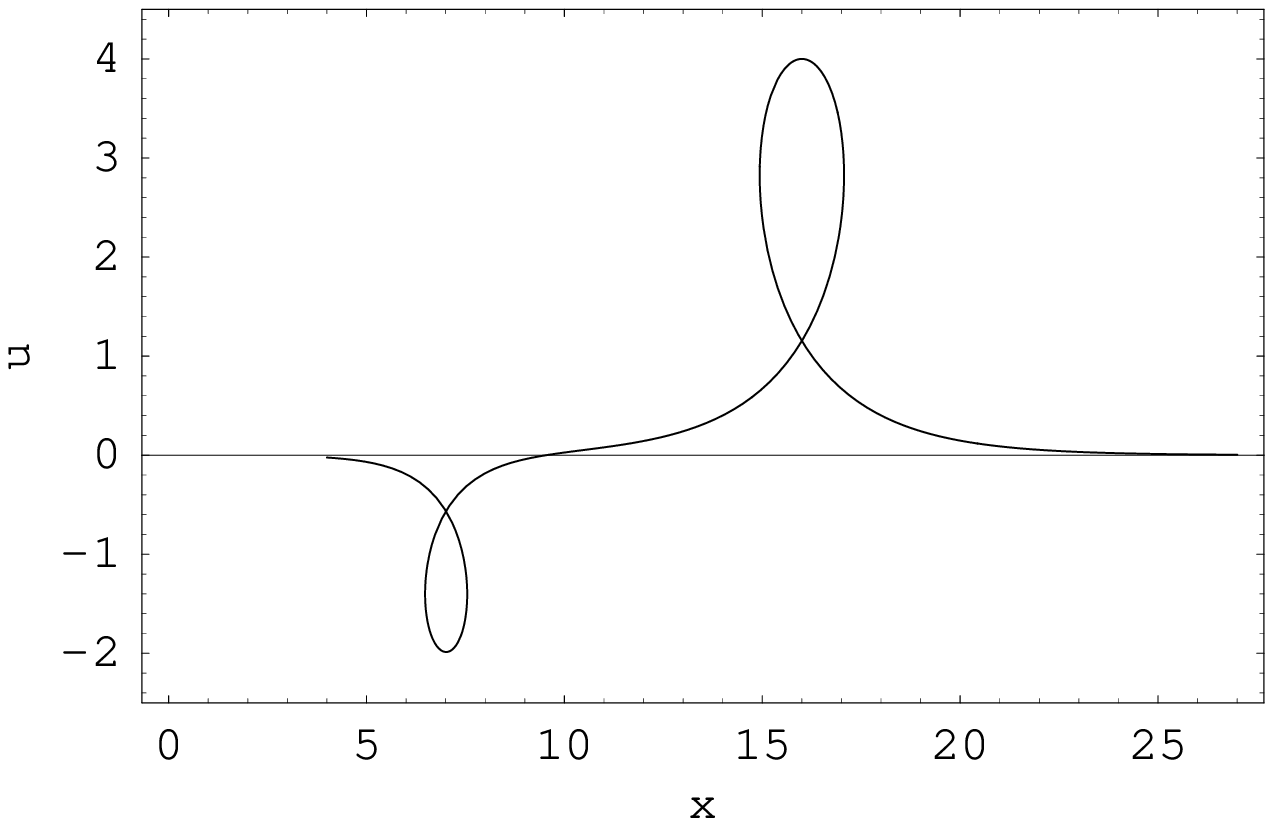}
\end{center}
\centerline{{\bf Fig. 6}: The profile of the loop-antiloop soliton solution.} \par
\bigskip
The formulas for the phase shift are given by
$$\Delta_1={1\over p_1}\ln\left({p_1-p_2\over p_1+p_2}\right)^2+{4\over p_2},\eqno(4.21a)$$
$$\Delta_2={1\over p_2}\ln\left({p_1-p_2\over p_1+p_2}\right)^2+{4\over p_1}.\eqno(4.21b)$$
Notice that the formula for $\Delta_1$ is altered when compared with (4.20a). 
Figure 5 shows the interaction of a loop soliton and an antiloop soliton with the parameters given by $p_1=0.5, p_2=-1.0,
c=\xi_{10}=\xi_{20}=0$ and Figure 6 shows the profile of the solution at $t=-5$. \par
\medskip
\leftline{\bf 4.3 Breather solutions}\par
\leftline{\bf 4.3.1 1-breather solution}\par
The breather solution of the sG equation is the bound state of the kink and antikink solutions.
Under certain condition, the breather solution is shown to
yield a nonsingular oscillating pulse solution of the SP equation,  which we shall term the
breather solution as well. The 1-breather solution of the SP equation is derived if we put
$N=2$ in (4.2) and specify the parameters as
$$p_1=a+{\rm i}b,\quad p_2=a-{\rm i}b, \eqno(4.22a)$$
$$\quad \xi_{10}=\lambda +{\rm i}\mu,
\quad \xi_{20}=\lambda -{\rm i}\mu, \eqno(4.22b)$$
where $a$ and $b$ are  positive constants and $\lambda$ and $\mu$ are real constants. Then, (4.2) gives
$$f=1+{\rm i}{\rm e}^{\xi_1}+{\rm i}{\rm e}^{\xi_1^*}+\left({b\over a}\right)^2{\rm e}^{\xi_1+\xi_1^*},\eqno(4.23a)$$
and $f^\prime=f^*$ where $\xi_1=\theta+{\rm i}\chi$ with
$$\theta=a\left(y+{1\over a^2+b^2}t\right)+\lambda, \eqno(4.23b)$$
$$\chi=b\left(y-{1\over a^2+b^2}t\right)+\mu. \eqno(4.23c)$$
Substituting (4.23) into (4.5) and (4.6), we obtain the following parametric representation of the solution
$$u(y,t)={4ab\over a^2+b^2}{b\sin \chi \cosh\left(\theta+\ln{b\over a}\right)-a \cos \chi \sinh\left(\theta+\ln{b\over a}\right)
\over b^2\cosh^2\left(\theta+\ln{b\over a}\right)+a^2\cos^2 \chi}, \eqno(4.24a)$$
$$x(y,t)=y-{2ab\over a^2+b^2}{a\sin 2\chi+b\sinh 2\left(\theta+\ln{b\over a}\right)
\over b^2\cosh^2\left(\theta+\ln{b\over a}\right)+a^2\cos^2 \chi}
-{4a\over a^2+b^2}+c. \eqno(4.24b)$$
Note that $u$ has two different phase variables $\theta$ and $\chi$.
The phase $\theta$ characterizes the envelope of the breather whereas the phase $\chi$ governs
the internal oscillation. In general, the solution (4.24) would exhibit singularities. 
Unlike the loop soliton solutions, the solution becomes a nonsingular function of $x$ and $t$
if we impose a condition for the parameters $a$ and $b$. To see this, we apply the criterion (4.8) to
the $\tau$-function (4.2) to obtain 
$$-\sqrt{2}+1<{a\over b}{\cos \chi\over \cosh\left(\theta+\ln{b\over a}\right)}<\sqrt{2}-1. \eqno(4.25a)$$
This inequality must be satisfied for any value of $\theta$ and $\chi$.
Since $a>0$ and $b>0$, we see from (4.25) that the condition imposed on the parameters turns out to be
$$0<a/b<\sqrt{2}-1. \eqno(4.25b)$$
  \begin{center}
\includegraphics[width=10cm]{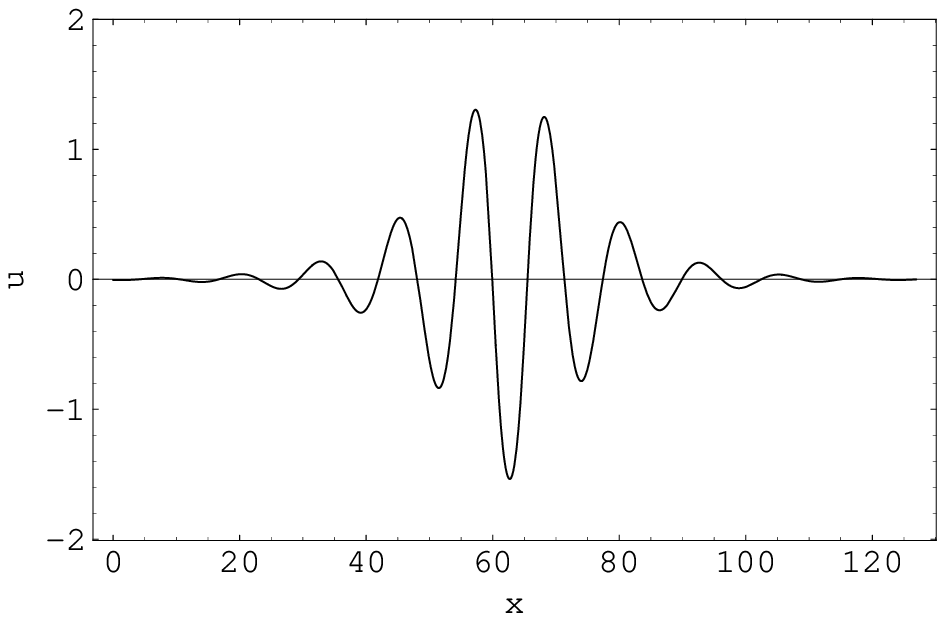}
\end{center}
\centerline{{\bf Fig. 7}: The profile of the nonsingular 1-breather solution.} \par
\bigskip
Figure 7 shows a profile of the 1-breather solution at $t=0$ with the 
parameters $a=0.1, b=0.5, c=80, \lambda=\mu=0$. In this example, $a/b=0.2$ so that there appear no
singularities as expected from the criterion (4.25b). 
For completeness, it will be instructive to present a singular breather solution. Figure 8 shows an
example of the singular solution with the parameters $a=0.4, b=0.5, c=20, \lambda=\mu=0$. Obviously, the criterion
for the nonsingular solution is violated.
\par
\begin{center}
\includegraphics[width=10cm]{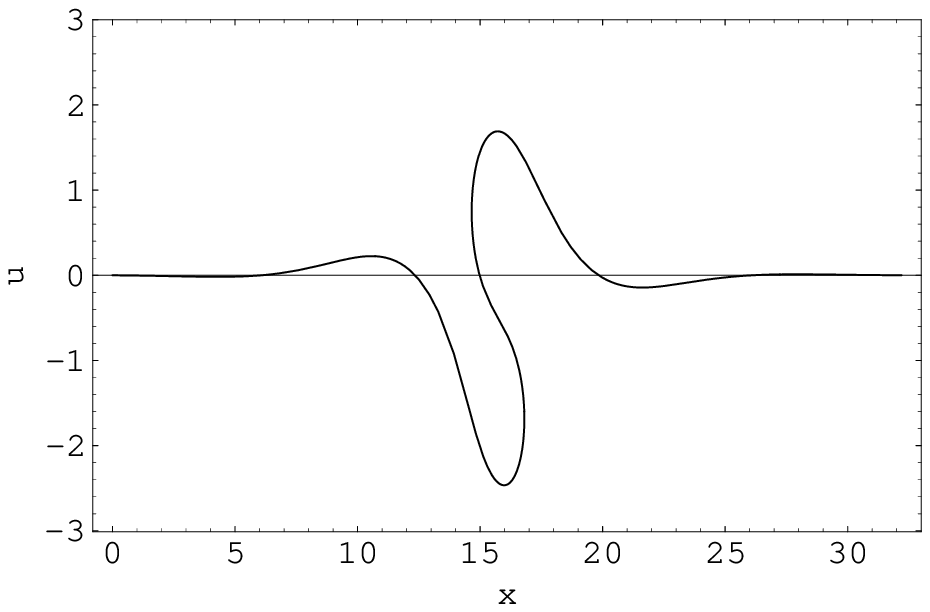}
\end{center}
\centerline{{\bf Fig. 8}: The profile of the singular 1-breather solution.} \par
\bigskip
\leftline{\bf 4.3.2 M-breather solution}\par
The general $M$-breather solution is constructed from the $M$-breather solution of the sG
equation (4.1) and (4.2) with $N=2M$. Specifically, we set
$$p_{2j-1}=p_{2j}^*\equiv a_j+{\rm i}b_j,\quad a_j>0,\quad b_j>0,
\quad (j=1, 2, ..., M),\eqno(4.26a)$$
$$\xi_{2j-1,0}=\xi_{2j,0}^*\equiv \lambda_j+{\rm i}\mu_j,\quad (j=1, 2, ..., M),\eqno(4.26b)$$
and write  the phase variables $\xi_{2j-1}$ and $\xi_{2j}$  as
$$\xi_{2j-1}=\theta_j+{\rm i}\chi_j, \quad (j=1, 2, ..., M),\eqno(4.27a)$$
$$\xi_{2j}=\theta_j-{\rm i}\chi_j, \quad (j=1, 2, ..., M),\eqno(4.27b)$$
with
$$\theta_j=a_j(y+c_jt)+\lambda_j, \quad (j=1, 2, ..., M), \eqno(4.27c)$$
$$\chi_j=b_j(y-c_jt)+\mu_j, \quad (j=1, 2, ..., M), \eqno(4.27d)$$
$$c_j={1\over a_j^2+b_j^2}, \quad (j=1, 2, ..., M). \eqno(4.27e)$$
The parametric solution (4.5) and (4.6) with (4.26) and (4.27) describes
multiple collisions of $M$ breathers provided that certain condition is
imposed on the parameters $a_j$ and $b_j (j=1, 2, ..., M)$. In the present
$M$-breather case, the simple inequality like (4.25b) is still difficult to
obtain. However, as shown below,  the $M$-breather solution splits into $M$
single breathers as $t\rightarrow \pm\infty$. Hence, one can expect that the
condition corresponding to (4.25b) would become
$$0<\sum_{j=1}^M{a_j\over b_j}<\sqrt{2}-1. \eqno(4.28)$$
It will be demonstrated later that the 2-breather solution exists whose parameters 
indeed satisfy the inequality (4.28).
\par
Let us now investigate the structure of the $M$-breather solution by focusing
on the asymptotic behavior for large time. To this end,
 we order the magnitude of the velocity of each breather as
$c_1>c_2>... >c_M$. We take the limit $t\rightarrow -\infty$ with $\theta_n$
being fixed. Then, we see that
 $u$ and $x$ have the leading-order asymptotics
$$u(y,t)\sim {4a_nb_n\over a_n^2+b_n^2}{G_n
\over F_n}, \eqno(4.29a)$$
$$x(y,t)\sim y-{2a_nb_n\over a_n^2+b_n^2}{H_n
\over  F_n}   -{4a_n\over a_n^2+b_n^2}+d, \eqno(4.29b)$$
  with
  $$F_n=b_n^2\cosh^2\left(\theta_n+\alpha_n^{(-)} +\ln{b_n\over a_n}\right)+a_n^2\cos^2 \left(\chi_n+\beta_n^{(-)}\right),\eqno(4.29c)$$
    $$G_n=b_n\sin\left(\chi_n+\beta_n^{(-)}\right)
 \cosh\left(\theta_n+\alpha_n^{(-)}+\ln{b_n\over a_n}\right)$$
 $$-a_n \cos\left(\chi_n+\beta_n^{(-)}\right)
  \sinh\left(\theta_n+\alpha_n^{(-)}+\ln{b_n\over a_n}\right),\eqno(4.29d)$$
  $$H_n=a_n\sin 2\left(\chi_n+\beta_n^{(-)}\right)
  +b_n\sinh 2\left(\theta_n+\alpha_n^{(-)} +\ln{b_n\over a_n}\right),\eqno(4.29e)$$
  where the real parameters $\alpha_n^{(-)}$ and $\beta_n^{(-)}$ are defined by the relation
$$\sum_{j=2n+1}^{2M}\ln\left({p_{2n-1}-p_j\over p_{2n-1}+p_j}\right)^2=\alpha_n^{(-)}+{\rm i}\beta_n^{(-)}.\eqno(4.30a)$$
The explicit expressions of $\alpha_n^{(-)}$ and $\beta_n^{(-)}$ in terms of $a_j$ and $b_j$ 
are calculated using (4.26a). They read
$$\alpha_n^{(-)}=\sum_{j=n+1}^{M}\ln{\{(a_n-a_j)^2+(b_n-b_j)^2\}\{(a_n-a_j)^2+(b_n+b_j)^2\}
\over \{(a_n+a_j)^2+(b_n+b_j)^2\}\{(a_n+a_j)^2+(b_n-b_j)^2\}},\eqno(4.30b)$$
$$\beta_n^{(-)}=2\sum_{j=n+1}^M\Biggl(\tan^{-1}{b_n-b_j\over a_n-a_j}+\tan^{-1}{b_n+b_j\over a_n-a_j}$$
$$-\tan^{-1}{b_n+b_j\over a_n+a_j}-\tan^{-1}{b_n-b_j\over a_n+a_j}\Biggr).\eqno(4.30c)$$
\par
As $t \rightarrow +\infty$, $u$ and $x$ take the same asymptotic forms as (4.29) with
 $\alpha_n^{(-)}$ and $\beta_n^{(-)}$ replaced by $\alpha_n^{(+)}$ and $\beta_n^{(+)}$, respectively where
 $$\sum_{j=1}^{2n-2}\ln\left({p_{2n-1}-p_j\over p_{2n-1}+p_j}\right)^2=\alpha_n^{(+)}+{\rm i}\beta_n^{(+)}.\eqno(4.31)$$
The expressions of $\alpha_n^{(+)}$ and $\beta_n^{(+)}$ corresponding to (4.30b,c) follow if one replaces the
sum $\sum_{j=n+1}^M$ by $\sum_{j=1}^{n-1}$ in (4.30b,c). Observing the asymptotic behavior of the solution
in the rest frame of reference, we see that it represents a superposition of $M$ breathers, each has a form given by
(4.24a). The  effect of the interaction is the phase shift 
 given by the sum of the quantities
$\alpha_n^{(\pm)}$ and $\beta_n^{(\pm)}$  which is caused by
 the pair wise collisions of $M$ breathers and a term due to the coordinate transformation. \par
The formula for the total phase shift will not be defined definitely since the solution
takes the form of wave packets. If we consider the small-amplitude limit, however, the oscillating part and
the envelope are shown to be separated completely so that one can obtain the formula like (4.17)
for the $N$-loop soliton solution. Actually, in the small-amplitude limit $a_n \rightarrow 0\ (n=1, 2, ..., M)$,
 expressions (4.29) and (4.30) are approximated by
$$u(y,t) \sim {4a_n\over b_n^2}{\sin\left(\chi_n+\beta_n^{(-)}\right) \over
\cosh \left(\theta_n+\alpha_n^{(-)}+\ln{b_n\over a_n}\right)}, \eqno(4.32a)$$
$$x(y,t) \sim y-{4a_n\over b_n^2}\tanh \left(\theta_n+\alpha_n^{(-)}+\ln{b_n\over a_n}\right) -{4a_n\over b_n^2}+c,
\eqno(4.32b)$$
$$\alpha_n^{(-)} \sim -\sum_{j=n+1}^M{8(b_n^2+b_j^2)a_ja_n\over (b_n^2-b_j^2)^2}, \eqno(4.33a)$$
$$\beta_n^{(-)} \sim \sum_{j=n+1}^M{8a_jb_n\over b_n^2-b_j^2}. \eqno(4.33b)$$
Let $\bar\Delta_n$ be the phase shift of the center position ( a point corresponding to the maximum amplitude) 
of the envelope for the
$n$th breather. By performing the asymptotic analysis similar to that  for the $N$-loop soliton case, we
find that
$$\bar\Delta_n=\sum_{j=n+1}^M{8(b_n^2+b_j^2)a_j\over (b_n^2-b_j^2)^2}
   -\sum_{j=1}^{n-1}{8(b_n^2+b_j^2)a_j\over (b_n^2-b_j^2)^2}, \quad (n=1, 2, ..., M). \eqno(4.34)$$
   \medskip
   \leftline{\bf 4.3.3  2-breather solution}\par
The solution describing the interaction of two breathers is parameterized 
by (4.26) and (4.27) with $M=2$. Since the parametric solution has a lengthy expression,
   it is not appropriate to write it down here. Instead, 
    we show the 
   time evolution of the solution graphically and demonstrate its
   solitonic behavior. Fig. 9 depicts the profile of the two-breather solution for three different
   times $(a) t=-40, (b) t=-5, (c) t=35$. The values of the parameters are chosen as $a_1=0.1, b_1=0.5, a_2=0.16,
   b_2=0.8, \lambda_1=\lambda_2=0, \mu_1=\mu_2=0$. The velocity of the large breather is $3.85$ while
   that of the small breather is $1.50$ (see (4.27e)).
   Note in this example $\sum_{j=1}^2(a_j/b_j)=0.4$ so that the inequality (4.28) is
   satisfied. For large negative time, the solution behaves like two independent breathers, each has
   a form given by (4.24) and propagates to the left.   As time goes, both breathers merge and then they
   separate each other with leaving the original wave profiles. Apparently, the present example 
   exhibits a typical feature common to    the interaction of  two-soliton    solutions. \par
      \medskip
   (a) \par
   \begin{center}
\includegraphics[width=10cm]{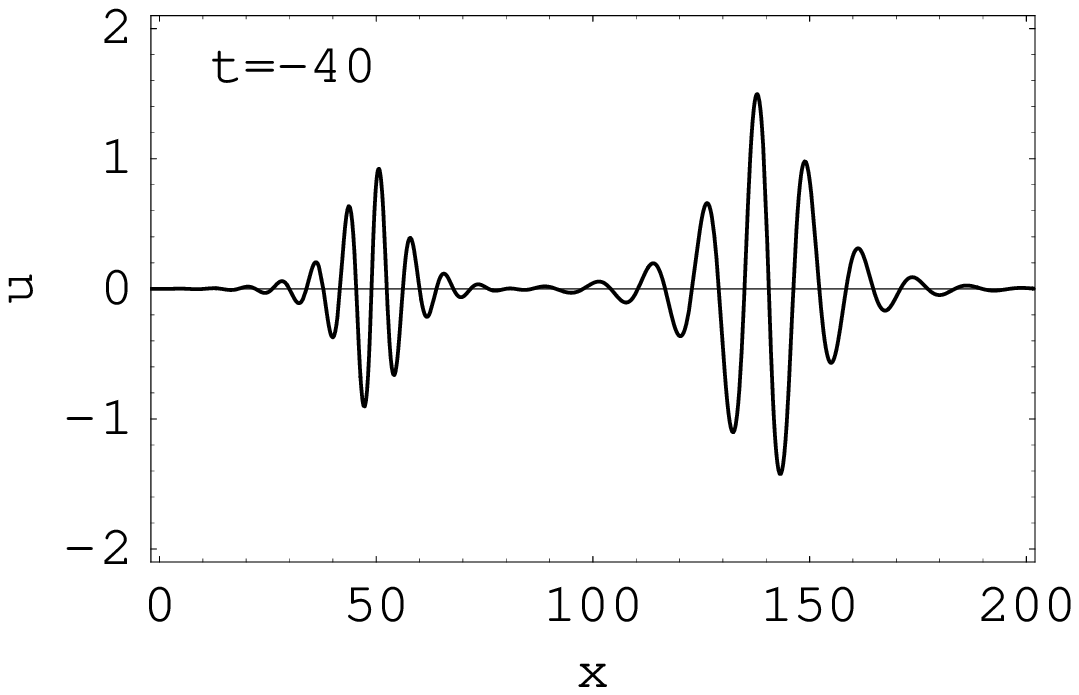}
\end{center}
\medskip
\newpage
(b) \par
   \begin{center}
\includegraphics[width=10cm]{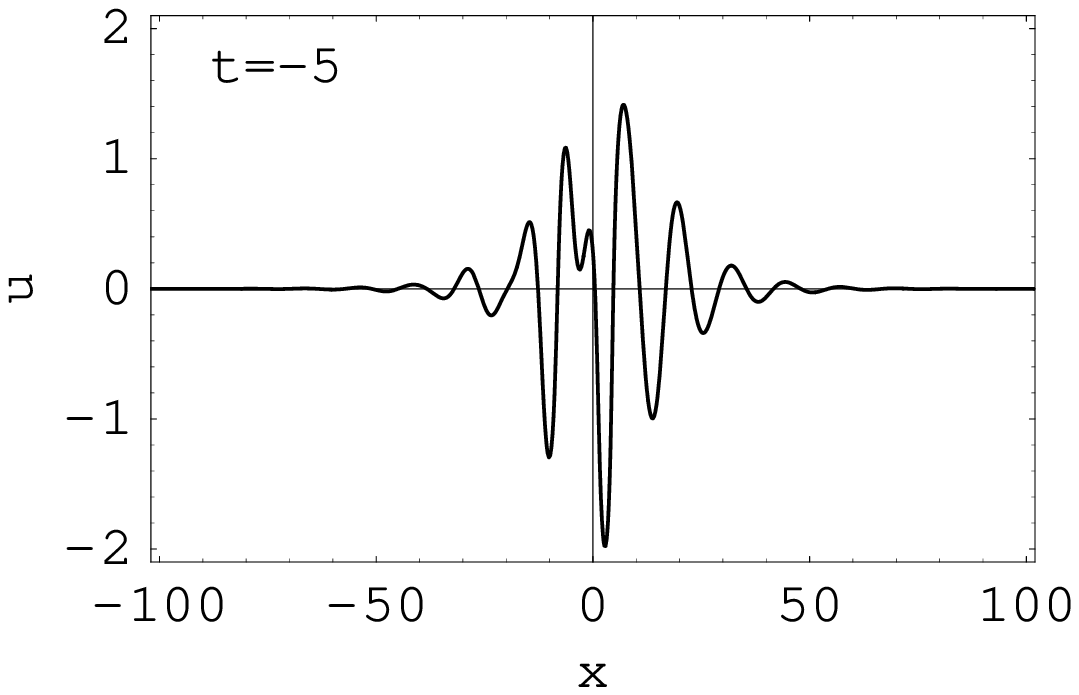}
\end{center}
\par
 (c) \par
   \begin{center}
\includegraphics[width=10cm]{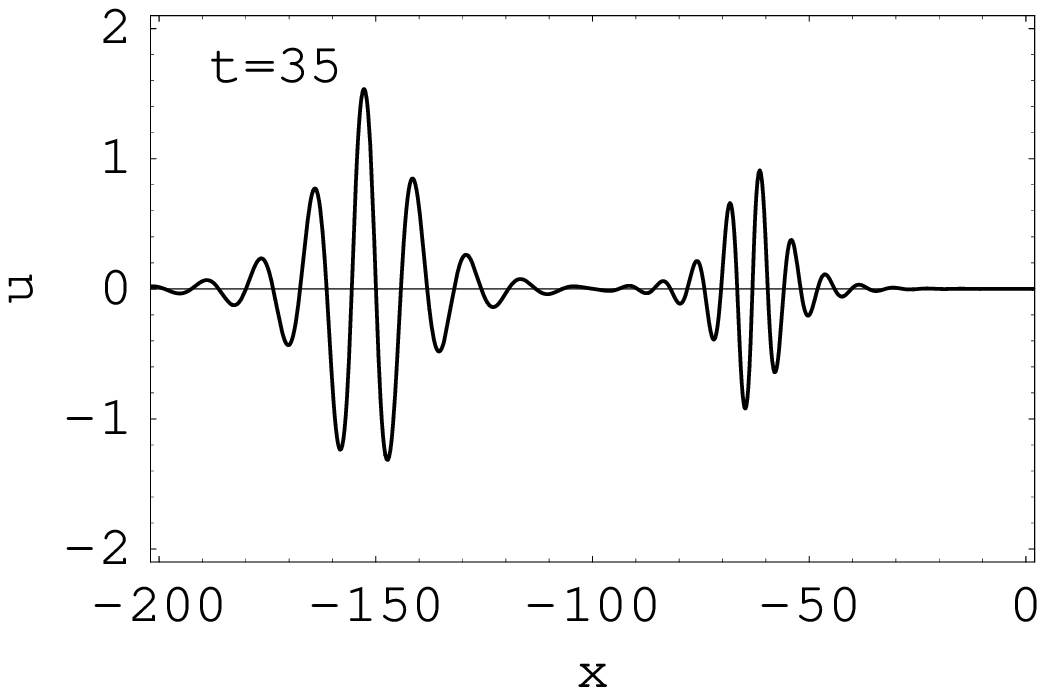}
\end{center}
\noindent {\bf Fig. 9}: The profile of the 2-breather solution for (a) $t=-40$, (b) $t=-5$, and (c) $t=35$. \par
   \bigskip
   \leftline{\bf 4.4 Remark}\par
   The 1- and 2-loop soliton solutions as well as the 1-breather solution have been obtained
   by different methods [14, 15]. \par
   \bigskip
\leftline{\bf 5 PERIODIC SOLUTIONS}\par
\leftline{\bf 5.1 1-phase solutions} \par
The periodic solutions of the SP equation can be constructed by using an exact method of solution described in
Sec. 3 [16]. Here, we deal with solutions which depend on the single variable $\eta$. 
    Then, the sG equation (3.11) reduces to an ODE for $\phi$
$$\phi^{\prime\prime}= -\sin \phi, \eqno(5.1)$$
where the prime appended to $\phi$ denotes the differentiation with respect to $\eta$.
    There exist several particular solutions of Eq. (5.1).  Among them, we look for solutions expressed
    by Jacobi's elliptic functions. Correspondingly, these solutions yield the 1-phase solutions of the
    SP equation. We investigate the properties of the solutions by the following two  examples. \par
    \medskip
\noindent{\bf 5.1.1 Example 1} \par
The first example of the solution of Eq. (5.1) is given by Jacobi's ${\rm sn}$ function. Explicitly
$$\phi=-2 \sin^{-1}{\rm sn}\left({\eta\over k}, k\right), \eqno(5.2)$$
  where the parameter $k$ is the modulus of the elliptic function. Substituting (5.2) into (3.12) gives
  the parametric representation of $u$
  $$u={2\over ka}{\rm dn}\left({\eta\over k}, k\right), \eqno(5.3)$$
where ${\rm dn}(u,k)$ is Jacobi's ${\rm dn}$ function. If we introduce the relation $\cos \phi=1-2\,{\rm sn}^2\left({\eta\over k}, k\right)$
which is derived from (5.2) into (3.14) , we obtain 

    $$x=y-2\int{\rm sn}^2\left({\eta\over k}, k\right)dy + c. \eqno(5.4)$$
    One can see that the integration constant depends on $t$ whose time evolution is determined by the second equation of
    (3.13) with $u$ given by (5.3). Indeed, using the identity
$$ k^2\,{\rm sn}^2\left({\eta\over k}, k\right)+{\rm dn}^2\left({\eta\over k}, k\right)=1, \eqno(5.5)$$
we obtain an ODE for $c$, $c^\prime(t)=-2/(ka)^2$. This equation can be integrated immediately to give
$$c(t)=-{2\over (ka)^2}t +d, \eqno(5.6)$$
where $d$ is an integration constant. Substituting (5.5) and (5.6) into (5.4) and rearranging terms, we find a
parametric representation of $x$
$$x+{2-k^2\over (ak)^2}t - x_0={1\over ak}\left\{-{1\over k}(2-k^2)\eta +2E\left({\eta\over k}, k\right)\right\}+d,
\eqno(5.7)$$
where $x_0=-\eta_0/a$ and $E(u, k)$ is the elliptic integral of the second kind defined by [17]
$$E(u, k)=\int^u_0{\rm dn}^2v\ dv=\int^\tau_0\sqrt{1-k^2t^2\over 1-t^2}dt,\ (t={\rm sn}\ v, \tau= {\rm sn}\ u). \eqno(5.8)$$
    Thus, (5.3) and (5.7) give the parametric solution of the 1-phase solution. This solution becomes a multi-valued
    function. In fact, applying the criterion (3.15) to (5.2), we see that $u_x$ exhibits singularity when ${\rm sn}(\eta/k, k)=\pm 1/\sqrt{2}$.
    To investigate the properties of the solution, it is convenient to take 
    the amplitude and the
modulus as independent parameters. The amplitude $A$ of the wave may be defined by the relation
$A=(u_{\rm max}-u_{\rm min})/2$ where $u_{\rm max}$ and $u_{\rm min}$ are the maximum and minimum values of $u$, respectively.
In the present example, it follows from (5.3) that 
$$A={1-\sqrt{1-k^2}\over ka}, \eqno(5.9)$$
which enables us to express the parameter $a$ in terms of $A$ and $k$. The velocity $V$ and the wavelength $\Lambda$
of the wave are then given respectively by
$$V={2-k^2\over (ak)^2}={(2-k^2)A^2\over (1-\sqrt{1-k^2})^2}, \eqno(5.10)$$ 
$$  \Lambda={2A\over 1-\sqrt{1-k^2}}\left\{(2-k^2)K(k)-2E(k)\right\}, \eqno(5.11)$$
where $K(k)$ and $E(k)$ are the complete elliptic integrals of the first and second kinds, respectively. 
In deriving (5.11), we have used (5.7) and the periodicity relation 
$$E(u+2K(k), k)=E(u, k)+2E(k). \eqno(5.12)$$
    The wavenumber $K$ (which should not be confused with $K(k)$) and the angular frequency $W$ are defined respectively by
$K=2\pi/\Lambda$ and $W=VK$. The limiting forms of these wave parameters for  $k\rightarrow 0$ and
 $k\rightarrow 1$ are given respectively by
 $$K\sim {8\over Ak^2}, \ V\sim {8A^2\over k^4}, \ W\sim {64A^2\over k^6}, \  (k\rightarrow 0), \eqno(5.13a)$$
$$K\sim{2\pi\over A}{1\over {\rm ln}{8\over 1-k}}, \ V\sim A^2, \ W\sim {2\pi A\over {\rm ln}{8\over 1-k}}, \ (k\rightarrow 1).\eqno(5.13b)$$
It follows from (5.13)  that 
$$W\sim A^2K\ (K\rightarrow 0), \eqno(5.14a)$$
$$ W\sim A^5K^3/8\ (K\rightarrow \infty). \eqno(5.14b)$$
    \begin{center}
\includegraphics[width=10cm]{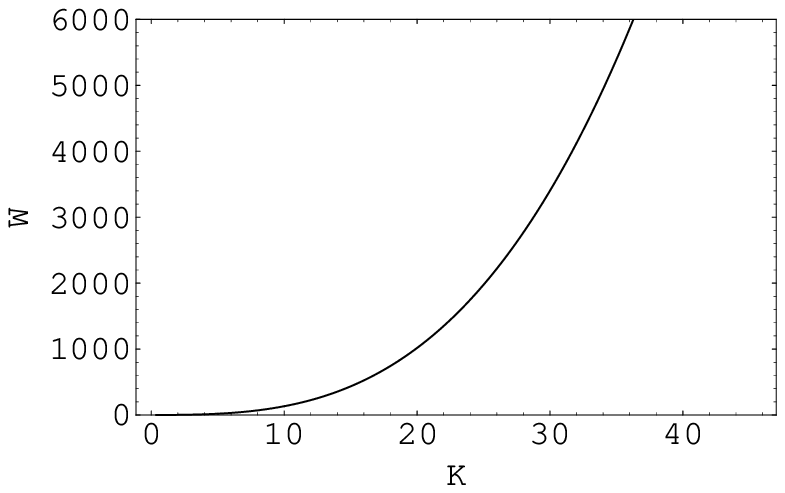}
\end{center}
\noindent {\bf Fig. 10}: The dispersion relation $W=W(K, A)$ with $A=1.0$ for Example 1 as a function of $K$.\par
\bigskip
\begin{center}
\includegraphics[width=10cm]{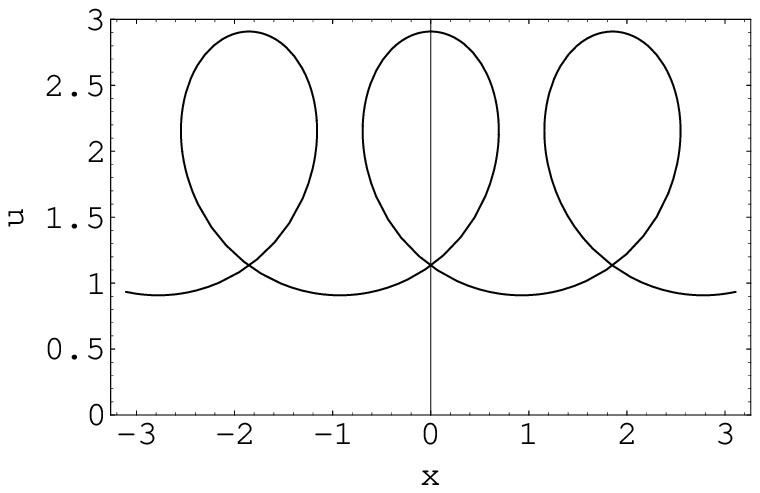}
\end{center}
\noindent {\bf Fig. 11}: A typical profile of the periodic solution for Example 1. \par
\bigskip
 The dispersion relation $W=W(K, A)$  with $A=1.0$ is plotted in Fig. 10 as a function of $K$. 
 Figure 11 illustrates the typical profile of the periodic solution  represented by (5.3) and (5.7) where the parameters are
chosen as $A=1.0, k=0.95, x_0=d=\eta_0=0$ and the time $t$ is set to zero. In this example, one can
see that the period $\Lambda$ is 1.853.  The figure represents a periodic loop traveling to the left
at a constant velocity $V=2.320$.\par
When the wavelength of the periodic wave becomes very long, it degenerates into a single loop soliton,
as we shall now demonstrate.
 As seen from (5.13), the long-wave limit $\Lambda\rightarrow \infty$ (or $K\rightarrow 0$)  is attained when $k$ tends to 1. 
Using the relations ${\rm dn}(u, 1)= {\rm sech}\ u$ and $E(u, 1)=\tanh u$, the parametric solution  
represented by (5.3) and (5.7) reduces respectively to
$$u=2A\ {\rm sech}\ \eta, \eqno(5.15a)$$
$$x+A^2t+x_0=-A\eta+2A\tanh\ \eta +d, \eqno(5.15b)$$
with $\eta=y/A-At+\eta_0$. We see from this expression that the limiting solution is essentially the same as that of the 1-loop
soliton solution given by (4.10a) and (4.11).  See Fig. 1. \par
\medskip
\noindent{\bf 5.1.2 Example 2} \par
    The second example of the solution of Eq. (5.1) is given by Jacobi's ${\rm dn}$ function
$$\phi=-2 \cos^{-1}{\rm dn}(\eta, k). \eqno(5.16)$$
    The parametric representation of the solution  can be written in the form
$$u={2k\over a}\,{\rm cn}(\eta, k), \eqno(5.17a)$$
$$x-{1\over a^2}(1-2k^2)t-x_0={1\over a}\left\{-\eta+2E(\eta, k)\right\}+d, \eqno(5.17b)$$
where ${\rm cn}(\eta, k)$ is Jacobi's ${\rm cn}$ function. This solution is characterized by the
wave parameters
$$A={2k\over a},\eqno(5.18)$$
$$V={1\over a^2}(1-2k^2)={A^2\over 4k^2}(1-2k^2), \eqno(5.19)$$
$$\Lambda={2A\over k}|-K(k)+2E(k)|. \eqno(5.20)$$
     Figure 12 shows the dispersion
relation $W=W(K, A)$ with $A=1.0$. \par
   The dispersion curve has two branches depending on the value of $k$.
The upper branch plotted by the solid line corresponds to
the dispersion relation for $0\leq k\leq k_c$ whereas the lower one (broken line) represents the dispersion
relation for $k_c<k\leq 1$ where $k_c(=0.9089)$ is a solution of the transcendental equation $K(k)=2E(k).$ 
    Note that the wavelength $\Lambda$ becomes zero when $k=k_c$. \par
        \bigskip
     \begin{center}
\includegraphics[width=10cm]{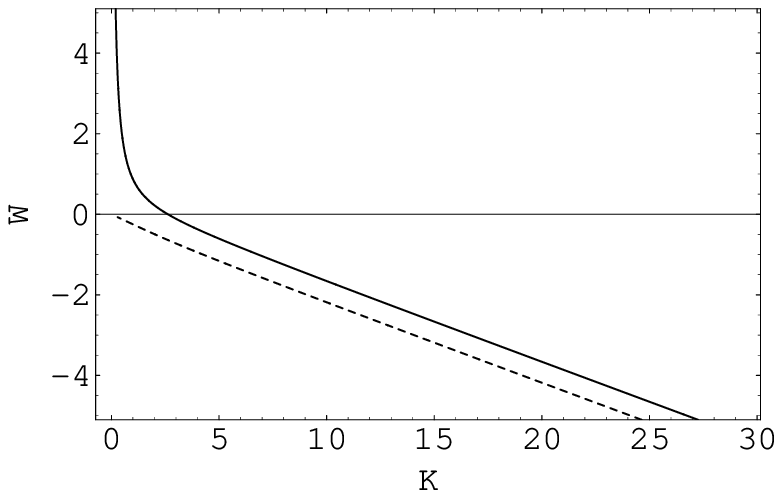}
\end{center}
\noindent {\bf Fig. 12}: The dispersion relation $W=W(K, A)$ with $A=1.0$ for Example 2 as a function of $K$.\par
\bigskip
    The limiting forms of the wave parameters for both $k\rightarrow 0$ and $k\rightarrow 1$ are given respectively by
$$ K\sim {2k\over A}, \ V\sim {A^2\over 4k^2}, \ W\sim {A\over 2k},\ (k\rightarrow 0). \eqno(5.21a)$$
$$K\sim{2\pi\over A}{1\over {\rm ln}{8\over 1-k}}, \ V\sim -{A^2\over 4}, 
\ W\sim -{\pi A\over 2\ {\rm ln}{8\over 1-k}},\ (k\rightarrow 1).\eqno(5.21b)$$
We see from (5.21) that $W\sim 1/K\ (K\rightarrow 0)$ for the upper branch and $K\sim -A^2K/4 \ (K\rightarrow 0)$
for the lower branch.  As $K\rightarrow \infty$, both branches approach a straight line $W=-V_cK$
with $V_c=(2k_c^2-1)A^2/(4k_c^2)\simeq 0.1974A^2$.  It is easy to see that the parametric solution (5.17) becomes
a single-valued function when   $k$ 
lies in the range $0<k<1/\sqrt{2}$. 
Note that the upper limit of this inequality coincides with the value of the modulus $k$ for which the
velocity given by (5.19) becomes zero. 
Figure 13 illustrates the  profile of the nonsingular  periodic solution at $t=0$ with  the parameters 
 $A=1.0, k=0.65, x_0=d=\eta_0=0$.  In this example, $\Lambda=3.027$. It represents a periodic wavetrain
 traveling to the right at a constant velocity $V=0.0917$. \par
 \bigskip
    \begin{center}
\includegraphics[width=10cm]{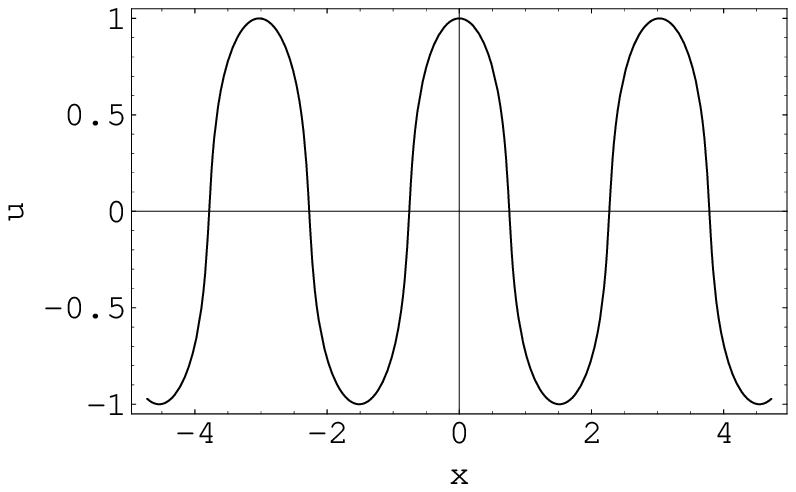}
\end{center}
\centerline{ {\bf Fig. 13}: A typical profile of the nonsingular periodic solution for Example 2.} \par
\bigskip
    If the parameter $k$ lies in the range $1/\sqrt{2}<k<1$, the solution exhibits singularities. 
    Figure 14 illustrates the profile of the nonsingular  periodic solution at $t=0$ with  the parameters 
 $A=1.0, k=0.8, x_0=d=\eta_0=0$.  In this example, $\Lambda=1.394$. It represents a periodic wavetrain
 traveling to the left at a constant velocity $V=0.1094$. \par
 \bigskip
    \begin{center}
\includegraphics[width=10cm]{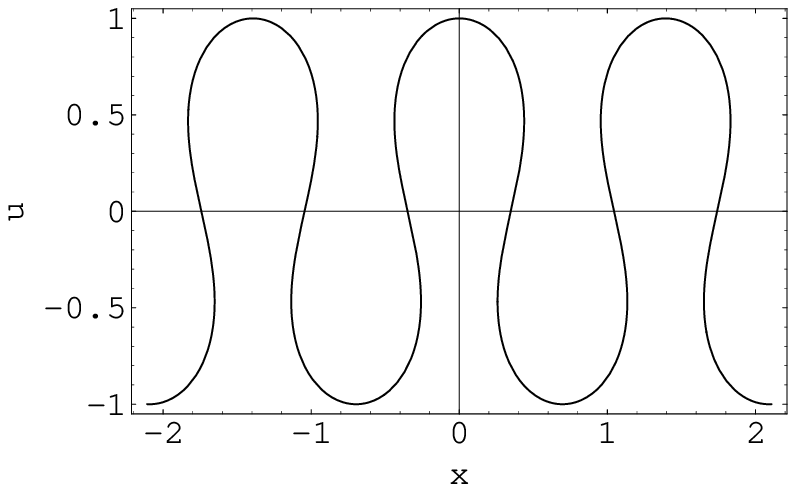}
\end{center}
\centerline{ {\bf Fig. 14}: A typical profile of the singular periodic solution for Example 2.} \par
\bigskip
    In conclusion, it will be worthwhile to consider the small amplitude limit of the solution. 
    As suggested by the asymptotic relations (5.21a), the appropriate limiting procedure is taken by the
    limit $k\rightarrow 0$ while keeping the values of $K, V$ and $W$ finite. It turns out that the
    magnitude of the amplitude $A$ is of order $k$.
    If we
substitute the relations ${\rm cn}(\eta, 0)=\cos\ \eta$ and $E(\eta, 0)=\eta$ into (5.17), the 
limiting form of the solution can be written as
$$u=A\ \cos\left(ax-{t\over a}-b\right), \eqno(5.22)$$
where $b=a(x_0+d)$ is a phase constant. The dispersion relation of this linear wave 
is given by $W=1/K$ as is consistent with the
asymptotics (5.21a). We also remark that (5.22) satisfies the linearized SP equation $u_{xt}=u$.
\par
\medskip
\leftline{\bf 5.2 2-phase solutions}\par
\leftline{\bf 5.2.1 Separation of variables} \par
The general $N$-phase solution is now available for the sG equation. See [18], for instance. However, it will be
difficult to perform the integral in (3.14) even for the 2-phase solution. 
An alternative approach for constructing the general $N$-phase solutions will be discussed in Sec. 6.
Here, we address
the following specific form introduced by Lamb [19, 20]
$$\phi=4\tan^{-1}\left[{f(\xi)\over g(\eta)}\right]. \eqno(5.23)$$
We substitute (5.23) into the sG equation (3.11) and see that the variables $\xi$ and $\eta$
can be separated 
if $f$ and $g$ satisfy the following nonlinear ODEs
$${f^\prime}^2=-\kappa f^4+\mu f^2+\nu, \eqno(5.24a)$$
$${g^\prime}^2=\kappa g^4+(\mu-1)g^2-\nu, \eqno(5.24b)$$
where $\kappa, \mu$ and $\nu$ are arbitrary constants.
For special choice of these parameters, one can obtain
 solutions for $f$ and $g$ which are expressed
in terms of elliptic functions.  \par
If we substitute (5.23) into (3.12), we immediately obtain the parametric representation of $u$
$$u={4\over a}{f^\prime g +fg^\prime\over f^2+g^2}. \eqno(5.25)$$
On the other hand, it follows from (5.23) by an elementary calculation using formulas of trigonometric 
functions that 
$$\cos\ \phi=1-{8f^2g^2\over (f^2+g^2)^2}. \eqno(5.26)$$
The right-hand side of (5.26) can be modified in such a way that the integral in (3.14) can be
performed analytically. To this end, we introduce the function $Y=Y(\xi, \eta)$
$$Y={c_1(f^2)^\prime + c_2(g^2)^\prime\over f^2+g^2}, \eqno(5.27)$$
where $c_1$ and $c_2$ are constants to be determined later.
Note that $Y$ depends on the variables $y$ and $t$ through the relation (3.10). 
Now, we differentiate $Y$ by $y$ and use (5.24) to simplify the resultant expression.
After some calculations, we obtain
$$Y_y={a\over (f^2+g^2)^2}\Big[-2\kappa(c_1f^6+3c_1f^4g^2-3c_2f^2g^4-c_2g^6)
  -4c_2f^2g^2$$
  $$+2(c_1+c_2)\left\{-2fgf^\prime g^\prime+2\mu f^2g^2-\nu(f^2-g^2)\right\}\Big]. \eqno(5.28)$$
We set $c_1+c_2=0$ and $c_1=-2/a$ to reduce (5.28) in the form
$$Y_y=4\kappa(f^2+g^2)-{8f^2g^2\over (f^2+g^2)^2}. \eqno(5.29)$$
Comparing (5.26) and (5.29), we find that
$$\cos\ \phi=1+Y_y-4\kappa(f^2+g^2). \eqno(5.30)$$
Finally, we substitute (5.30) into (3.14) and 
take account of (5.27) with $c_1=-c_2=-2/a$.
Then, the integration with respect to $y$ 
can be performed trivially to give
the expression of $x$ in terms of $f$ and $g$
$$x=y-{4\over a}{ff^\prime-gg^\prime\over f^2+g^2}-4\kappa\int(f^2+g^2)dy + c. \eqno(5.31)$$
The time dependence of $c$ can be determined by the second equation of (3.13) with $u$ and $x$ 
given respectively by (5.25) and (5.31). It turns out that $c^\prime(t)=0$ so that $c=d$(=const.).
The expressions (5.25) and (5.31) provide the parametric representation of the the two-phase
periodic solutions of the SP equation. \par
We can obtain several 2-phase periodic solutions depending on the choice of the functions
$f$ and $g$. Here, we exemplify three solutions which reduce, in the long-wave limit, to breather 
solution (Example 1), 2-loop soliton solution (Example 2)
and loop-antiloop soliton  solution (Example  3). \par
\medskip
\leftline{\bf 5.2.2 Example 1} \par
The first example of $f$ and $g$ assumes the form
$$f(\xi)=A\ {\rm cn}(\beta\xi, k_f), \ g(\eta)={1\over {\rm cn}(\Omega\eta, k_g)}, \eqno(5.32)$$
where $A$,  $\beta$ and $\Omega$ are positive parameters and $k_f$ and $k_g$ are moduli of the elliptic
function. If we substitute (5.32) into (5.24), we can determine the parameters $\kappa$, $\mu$ and $\nu$
as well as $k_f$, $k_g$ and $\Omega$ in terms of $A$ and $\beta$. In particular,
$$k_f^2={A^2\over 1+A^2}\left(1+{1\over \beta^2(1+A^2)}\right), \eqno(5.33a)$$
$$k_g^2={A^2\over 1+A^2}\left(1-{1\over \Omega^2(1+A^2)}\right), \eqno(5.33b)$$
$$\Omega^2=\beta^2+{1-A^2\over 1+A^2}, \eqno(5.33c)$$
$$\kappa={\beta^2k_f^2\over A^2}, \ \mu=\beta^2(2k_f^2-1), \ \nu=\beta^2A^2(1-k_f^2). \eqno(5.33d)$$
Note from (5.33) and the inequality $0\leq k_f\leq 1$ that the parameter $\beta$ must be
restricted by the condition
$${A\over \sqrt{1+A^2}} \leq \beta, \eqno(5.34)$$
with arbitrary positive $A$. \par
Now, the parametric representation of $u$ follows from (3.12), (5.23) and (5.32). It reads
$$u={4A\over a}{-\beta\ {\rm sn}(\beta\xi, k_f){\rm dn}(\beta\xi, k_f) {\rm cn}(\Omega\eta, k_g)
+\Omega\ {\rm cn}(\beta\xi, k_f){\rm sn}(\Omega\eta, k_g){\rm dn}(\Omega\eta, k_g) \over
A^2{\rm cn}^2(\beta\xi, k_f){\rm cn}^2(\Omega\eta, k_g)+1}. \eqno(5.35)$$
The expression of $x$ is dervied by substituting (5.32) into (5.31) as
$$x=y+{4\beta\over a}{{\rm cn}(\beta\xi, k_f){\rm cn}(\Omega\eta, k_g)\over 
A^2{\rm cn}^2(\beta\xi, k_f){\rm cn}^2(\Omega\eta, k_g)+1}\Big\{A^2{\rm sn}(\beta\xi, k_f){\rm dn}(\beta\xi, k_f){\rm cn}(\Omega\eta, k_g)$$
$$-{\beta k_f^2\over \Omega {k_g^\prime}^2}\ {\rm cn}(\beta\xi, k_f){\rm sn}(\Omega\eta, k_g){\rm dn}(\Omega\eta, k_g)\Big\}$$
$$-{4\beta\over a}\left[E(\beta\xi, k_f)-{k_f^\prime}^2\beta\xi
-{\beta k_f^2\over A^2\Omega {k_g^\prime}^2}\left\{E(\Omega\eta, k_g)- {k_g^\prime}^2\Omega\eta\right\}\right] + d. \eqno(5.36)$$
Here we have used the following integral formulas for the Jacobi cn function in performing the integral in (5.51)
$$\int{\rm cn}^2(u, k)du={1\over k^2}\left\{E(u, k)-{k^\prime}^2u\right\}, \eqno(5.37a)$$
$$\int{1\over {\rm cn}^2(u, k)}du={1\over {k^\prime}^2}\left\{{{\rm sn}(u, k){\rm dn}(u, k)\over {\rm cn}(u, k)}
-E(u, k)+{k^\prime}^2u\right\}, \eqno(5.37b)$$
where $k^\prime=\sqrt{1-k^2}$. 
Note that $\int(f^2+g^2)dy=a^{-1}\int f^2(\xi)d\xi+a^{-1}\int g^2(\eta)d\eta$ by virtue of (3.10).
\par
Let us now describe some properties of the parametric solution given by (5.35) and (5.36). 
In general, $u$ is a multiply periodic function of $x$ for fixed $t$. Under certain
condition, however, it becomes a simply periodic function. To see this, we define the
two parameters $L_\xi$ and $L_\eta$ by $L_\xi\equiv 4K(k_f)/\beta$ and $L_\eta\equiv 4K(k_g)/\Omega$.
In view of the periodicity of Jacobi's elliptic
 functions like ${\rm sn}(u+4K(k), k)={\rm  sn}(u, k)$, $L_\xi(L_\eta)$ is the period of $u$
 with respect to $\xi(\eta)$. In accordance with the value of $A$, there arise two possible cases for the
 period of $u$. 
 When $0<A\leq A_c\ (A_c\simeq 2.1797)$,
 one can show with use of (5.33) that
 the inequality
$L_\eta<L_\xi$ holds for arbitrary positive values of $\beta$, and
at fixed $A$, both $L_\xi$ and  $L_\eta$ are monotonically decreasing functions of $\beta$ and
vanish as $\beta\rightarrow \infty$.
 If the ratio of both periods becomes a rational number, i.e., $L_\xi/L_\eta=m_\eta/m_\xi$
 with $(m_\xi, m_\eta)=1$ and $m_\xi<m_\eta$, then $u$ has a  period $L$ with
 respect to $y$ given by
 $$L={1\over a}m_\xi L_\xi={1\over a}m_\eta L_\eta. \eqno(5.38)$$
With use of the relation (5.38), the period $\Lambda$  with respect to $x$ is determined from (5.36). Indeed, using the 
periodicity of the elliptic integral of the second kind 
$E(u+2mK(k), k)=E(u, k) +2mE(k)\ (m: {\rm integer})$ as well as (5.33) and (5.38), the spatial period is found to be as
$$\Lambda=L\left[1-4\beta^2\left\{{E(k_f)\over K(k_f)}-{k_f^2\over A^2(1-k_g^2)}{E(k_g)\over K(k_g)}
+{1\over \beta^2(1+A^2)}\right\}\right]. \eqno(5.39)$$
\par
When $A_c<A$, on the other hand, the equation $L_\xi=L_\eta$ has a unique solution for $\beta$ and the corresponding expression
of the period is also given by (5.39).
\par
We remark that the solution presented here becomes a single-valued function when the parameter lies
in the rang $0<A<\sqrt{2}-1$. This  follows from (4.8) and (5.32) with the aid of the inequality
$|{\rm cn}(\beta\xi, k_f){\rm cn}(\Omega\eta, k_g)|\leq 1$.\par
Figure 15 depicts a profile of $u$ at $t=0$. The parameters chosen here are $A=0.2, m_\xi=1, m_\eta=2, a=1.0, 
d=\xi_0=\eta_0=0$. Solving (5.38) for $\beta$, one obtains $\beta=0.5832$ so that $\Omega=1.124, k_f=0.3837, k_g=0.0958, L=11.21$. 
Substituting these values into
(5.39), the period $\Lambda$ is found to be  $10.37$. \par
\bigskip
    \begin{center}
\includegraphics[width=10cm]{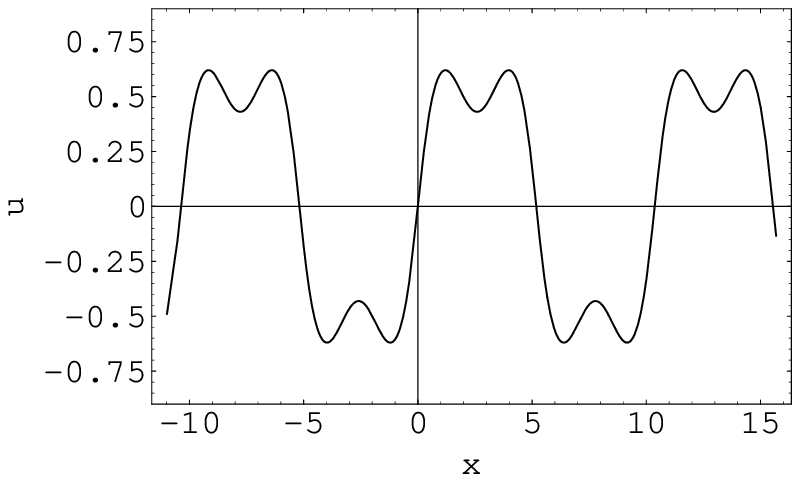}
\end{center}
\centerline{ {\bf Fig. 15}: A typical profile of the  periodic solution for Example 1.} \par
\bigskip
We  now consider the limiting profile  of the periodic solution when the period $\Lambda$ tends to
infinity.  
To be more specific, we take the limits $k_f\rightarrow 1$
and $k_g\rightarrow 0$. It then turns out from (5.33) that $\beta\rightarrow A/\sqrt{1+A^2}$ and
$\Omega \rightarrow 1/\sqrt{1+A^2}$. 
The limiting value of $\beta$ corresponds to the lower limit of the inequality (5.34).
Using the relations 
$${\rm sn}(u, 0)=\sin u, \ {\rm cn}(u, 0)=\cos u,
\ {\rm dn}(u, 0)=1, \eqno(5.40a)$$
$${\rm sn}(u, 1)=\tanh u,\  {\rm cn}(u, 1)={\rm sech}\ u, \ {\rm dn}(u, 1)={\rm sech}\ u, \eqno(5.40b)$$
$$E(u, 1)=\tanh u, E(u, 0)=u, \eqno(5.40c)$$
(5.35) and (5.36) reduces respectively to
$$u \sim {4A\Omega\over a}{\ -A\ \sinh \beta\xi\cos \Omega\eta+ \cosh \beta\xi\sin \Omega\eta 
\over \cosh^2\beta\xi+A^2\cos^2\Omega\eta }. \eqno(5.41a)$$
$$x \sim y-{2\Omega\over a}{\sinh 2\beta\xi+ A\ \sin 2\Omega\eta 
\over  \cosh^2\beta\xi+A^2\cos^2\Omega\eta} + d. \eqno(5.41b)$$
This parametric solution is essentially the same as the breather solution already given by (4.24). See Fig. 7. \par
\medskip
\leftline{\bf 5.2.3 Example 2} \par
The second example is given by the following $f$ and $g$ 
$$f(\xi)=A\ {{\rm sn}(\beta\xi, k_f)\over {\rm cn}(\beta\xi, k_f)}, \ g(\eta)={1\over {\rm dn}(\Omega\eta, k_g)}, \eqno(5.42)$$
where
$$k_f^2=1-A^2+{A^2\over \beta^2(1-A^2)}, \eqno(5.43a)$$
$$k_g^2=1-{1\over A^2}+{1\over \Omega^2(1-A^2)}, \eqno(5.43b)$$
$$\Omega=\beta A, \eqno(5.43c)$$
$$\kappa=-{\beta^2(1-k_f^2)\over A^2}, \ \mu=\beta^2(2-k_f^2), \ \nu=\beta^2A^2. \eqno(5.43d)$$
The inequalities $0\leq k_f\leq 1$ and $0\leq k_g\leq 1$
impose  the condition for $\beta$
$${1\over \sqrt{1-A^2}}\leq \beta\leq {1\over 1-A^2}, \eqno(5.44)$$
with $A$ in the range $0<A<1$. 
The exressions of $u$ and $x$ follows from (5.25), (5.31) and (5.42) with use of the formulas
$$\int{{\rm sn}^2(u, k)\over {\rm cn}^2(u, k)}du={1\over {k^\prime}^2}
\left({{\rm dn}(u, k){\rm sn}(u, k)\over {\rm cn}(u, k)}-\int {\rm dn}^2(u, k)du\right), \eqno(5.45a)$$
$$\int{1\over {\rm dn}^2(u, k)}du={1\over {k^\prime}^2}\left(-k^2\ {{\rm sn}(u, k){\rm cn}(u, k)
\over {\rm dn}(u, k)}+\int {\rm dn}^2(u, k)du\right). \eqno(5.45b)$$
The resulting parametric solution reads in the form \par
$$u={4A\over a}{\ \beta\ {\rm dn}(\beta\xi, k_f){\rm dn}(\Omega\eta, k_g)+k_g^2\Omega\ {\rm sn}(\beta\xi, k_f)
{\rm cn}(\beta\xi, k_f){\rm sn}(\Omega\eta, k_g){\rm cn}(\Omega\eta, k_g)
\over A^2{\rm sn}^2(\beta\xi, k_f){\rm dn}^2(\Omega\eta, k_g) + {\rm cn}^2(\beta\xi, k_f)}. \eqno(5.46)$$
$$x=y-{4\beta\over a}{1\over A^2{\rm sn}^2(\beta\xi, k_f){\rm dn}^2(\Omega\eta, k_g) + {\rm cn}^2(\beta\xi, k_f)}\times $$
$$\times\Big[(A^2{\rm dn}^2(\Omega\eta, k_g)-1){\rm sn}(\beta\xi, k_f){\rm cn}(\beta\xi, k_f){\rm dn}(\beta\xi, k_f)$$ 
$$+k_g^2A^3{\rm sn}^2(\beta\xi, k_f){\rm sn}(\Omega\eta, k_g){\rm cn}(\Omega\eta, k_g){\rm dn}(\Omega\eta, k_g)\Big]$$
$$+{4\beta\over a}\left(-E(\beta\xi, k_f) + A E(\Omega\eta, k_g)\right)+d. \eqno(5.47)$$
\par
Although the solution given above is a multiply periodic function, it has a single period if the condition
$$L={1\over 2a}m_\xi L_\xi={1\over 2a}m_\eta L_\eta, \eqno(5.48)$$
is satisfied.  Using the relation $(1-k_f^2)/{k_g^\prime}^2=A^4$ which follows from (5.43), the spatial period is
found to be as
$$\Lambda=L\left[1-4\beta^2\left\{{E(k_f)\over K(k_f)}-A^2{E(k_g)\over K(k_g)}\right\}\right]. \eqno(5.49)$$
Unlike Example 1, the solution always exhibit singularities as confirmed easily from (4.8) and (5.42).
\par
Figure 16 plots a profile of $u$ at $t=5$. The parameters are chosen as $A=0.2, m_\xi=2, m_\eta=1,
a=1.0, d=\xi_0=\eta_0=0$. In this example, $\beta=1.027, \Omega=0.2053, k_f=0.9998, k_g=0.8421,  L=20.35, \Lambda=5.938$. \par
\bigskip
    \begin{center}
\includegraphics[width=10cm]{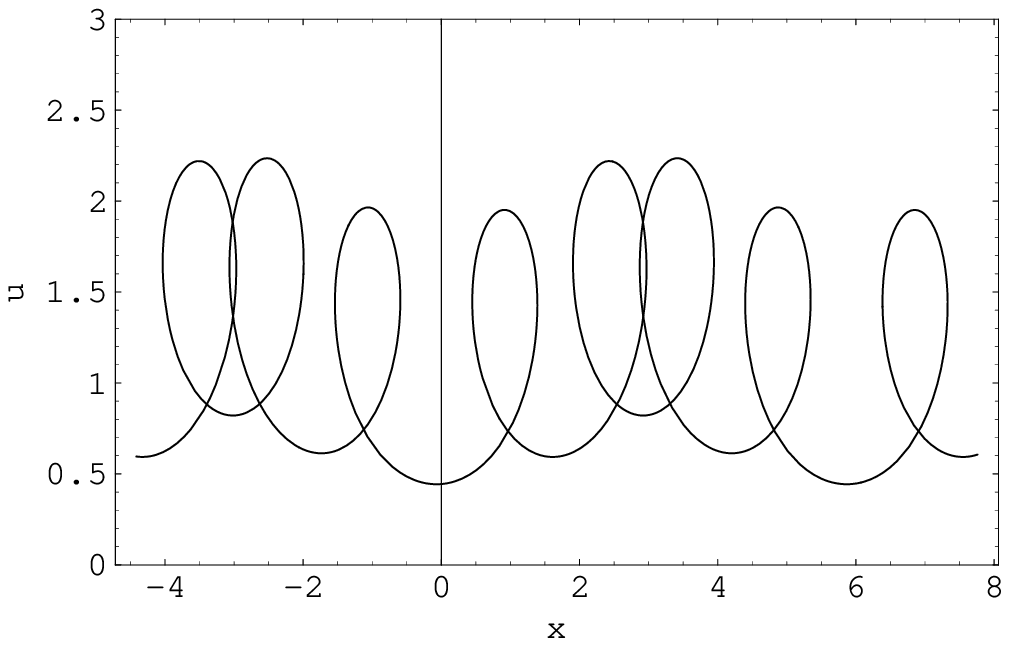}
\end{center}
\centerline{ {\bf Fig. 16}: A typical profile of the  periodic solution for Example 2.} \par
\bigskip
In considering the limiting profiles, there arise two cases according to the inequality (5.44). The upper
limit of the inequality for $\beta$ is attained when $k_g\rightarrow 0$. In this limit, one has the limiting forms
$$\Omega \sim {A\over 1-A^2}, \ k_f^2\sim 1-A^4, \ f\sim A{{\rm sn}(\beta\xi, k_f)
\over {\rm cn}(\beta\xi, k_f)},\ g\sim 1. \eqno(5.50)$$
 The expressions (5.46) and (5.47) then reduce respectively to
$$u \sim {4\over a}{\beta A\ {\rm dn}(\beta\xi, k_f)\over A^2{\rm sn}^2(\beta\xi, k_f)+{\rm cn}^2(\beta\xi, k_f)}
={4\over a}{\beta A(1+A^2)\ {\rm dn}(\beta\xi, k_f)\over {\rm dn}^2(\beta\xi, k_f)+A^2}, \eqno(5.51)$$
$$x \sim y+{4\over a}{(1+A^2) {\rm sn}(\beta\xi, k_f){\rm cn}(\beta\xi, k_f){\rm dn}(\beta\xi, k_f)
\over {\rm dn}^2(\beta\xi, k_f)+A^2}+{4\beta\over a}\left(-E(\beta\xi, k_f)+A\Omega\eta\right)+d, \eqno(5.52)$$
Using (3.10),  (5.52) is modified in the form
$$x+Vt-x_0={1\over a}(1+4\beta^2A^2)\xi+ {4\over a}{(1+A^2) {\rm sn}(\beta\xi, k_f){\rm cn}(\beta\xi, k_f){\rm dn}(\beta\xi, k_f)
\over {\rm dn}^2(\beta\xi, k_f)+A^2}$$
$$-{4\beta\over a}E(\beta\xi, k_f)+d, \eqno(5.53a)$$
with
$$V={1+6A^2+A^4\over[a(1-A^2)]^2}, \ x_0={1\over a}\left\{(1+4\beta^2A^2)\xi_0-4\beta^2A^2\eta_0\right\}. \eqno(5.53b)$$
    \begin{center}
\includegraphics[width=10cm]{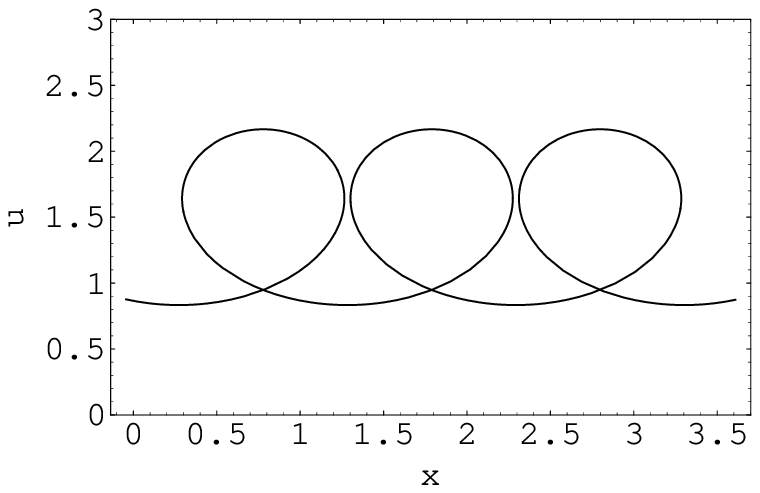}
\end{center}
\noindent {\bf Fig. 17}: Periodic loop arising from the limit $k_g\rightarrow 0$ of the periodic solution depicted in Fig. 16. \par
\bigskip
\noindent Since the parametric solution (5.51) and (5.53) depends only on the single variable $\xi$, it becomes a 1-phase
periodic function. Indeed,  as shown in Fig. 17, it represents a periodic train of loops propagating to the left at a constant velocity $V$.
See also Fig. 1 which illustrates a typical profile of a 1-loop soliton solution. 
The maximum and minimum values of $u$ are evaluated from (5.48).  They read as 
  $u_{\rm max}=2(1+A^2)/[a(1-A^2)]$ (at ${\rm dn}(\beta\xi,k_f)=A$) 
 and $u_{\rm min}=4A/[a(1-A^2)]$ (at ${\rm dn}(\beta\xi,k_f)=A^2, 1$). The spatial period $\Lambda$ 
 is given by (5.49) with $L=K(k_f)/(a\beta)$. In this example, $u_{\rm max}=2.167, u_{\rm min}=0.833, \beta=1.042,
 \Omega=0.2083, k_f=0.9992,  L=4.442, \Lambda=1.010, V=1.347$.
  \par
The lower limit of $\beta$ in (5.44) is realized when $k_f\rightarrow 1, \ k_g\rightarrow 1$, which leads to
 the asymptotics
 $$\Omega \sim {A\over \sqrt{1-A^2}},\ f\sim A\ \sinh \beta\xi, \ g\sim \cosh \Omega\eta. \eqno(5.54)$$
 Then, the parametric solution becomes
 $$u \sim {4\beta A\over a}{\cosh \beta\xi\cosh \Omega\eta +A \sinh \beta\xi \sinh \Omega\eta
 \over A^2 \sinh^2 \beta\xi +\cosh^2 \Omega\eta}, \eqno(5.55a)$$
 $$x \sim y -{2\beta\over a}{A^2 \sinh 2\beta\xi -A \sinh 2\Omega\eta \over 
A^2 \sinh^2 \beta\xi +\cosh^2 \Omega\eta} +d. \eqno(5.55b)$$
This expression coincides with the parametric form of the 2-loop soliton solution  given by (4.19). See Fig. 3. \par
\medskip
\leftline{\bf 5.2.4 Example 3} \par
The third example of $f$ and $g$ takes the form
$$f(\xi)=A\ {\rm dn}(\beta\xi, k_f),\ g(\eta)={{\rm cn}(\Omega\eta, k_g)\over {\rm sn}(\Omega\eta, k_g)}, \eqno(5.56)$$
where
$$k_f^2=1-{1\over A^2}+{1\over \beta^2(A^2-1)}, \eqno(5.57a)$$
$$k_g^2=1-A^2+{A^2\over \Omega^2(A^2-1)}, \eqno(5.57b)$$
$$\Omega={\beta\over A}, \eqno(5.57c)$$
$$\kappa={\beta^2\over A^2}, \ \mu=\beta^2(2-k_f^2), \ \nu=\beta^2A^2(k_f^2-1). \eqno(5.57d)$$
The inequalities  $0\leq k_f\leq 1$ and $0\leq k_g\leq 1$ require that the parameter $\beta$ always must lie in the range
$${A\over \sqrt{A^2-1}}\leq \beta\leq {A^2\over A^2-1}, \eqno(5.58)$$
with $A>1$.
 Using (5.25), (5.31) and (5.56), the parametric representation of the solution can be found to be as
 $$u=-{4A\over a}{\ \Omega\ {\rm dn}(\beta\xi, k_f){\rm dn}(\Omega\eta, k_g)+\beta k_f^2\ {\rm sn}(\beta\xi, k_f)
{\rm cn}(\beta\xi, k_f){\rm sn}(\Omega\eta, k_g){\rm cn}(\Omega\eta, k_g)
\over A^2{\rm dn}^2(\beta\xi, k_f){\rm sn}^2(\Omega\eta, k_g) + {\rm cn}^2(\Omega\eta, k_g)}, \eqno(5.59a)$$
$$x=y-{4\beta\over a}{1\over A^2{\rm dn}^2(\beta\xi, k_f){\rm sn}^2(\Omega\eta, k_g) + {\rm cn}^2(\Omega\eta, k_g)}\times$$
$$\times\Big[{1\over A}(1-A^2{\rm dn}^2(\beta\xi, k_f)){\rm sn}\ (\Omega\eta, k_g){\rm cn}\ (\Omega\eta, k_g){\rm dn}(\Omega\eta, k_g) $$
$$-k_f^2A^2{\rm sn}(\beta\xi, k_f){\rm cn}(\beta\xi, k_f){\rm dn}(\beta\xi, k_f){\rm sn}^2(\Omega\eta, k_g)\Big]$$
$$-{4\beta\over a}\left(E(\beta\xi, k_f) -{1\over A} E(\Omega\eta, k_g)\right)+d. \eqno(5.59b)$$
As demonstrated readily by using (4.8) and (5.56), this solution always becomes a multi-valued function. 
\par
A spatial period of the above solution can be found  if there exist integers $m_\xi$ and $m_\eta$ satisfying the
relation (5.48). Since in the present case  $K(k_f)/\beta<K(k_g)/\Omega$, one must impose the
condition $m_\eta<m_\xi, (m_\xi, m_\eta)=1$. The expression of the spatial period is now given by
$$\Lambda=L\left[1-4\beta^2\left\{{E(k_f)\over K(k_f)}-{1\over A^2}{E(k_g)\over K(k_g)}\right\}\right]. \eqno(5.60)$$
Figure 18 plots the profile of $u$ at $t=5$. The parameters are chosen as $A=5, m_\xi=2, m_\eta=1,
a=1.0, d=12, \xi_0=\eta_0=0$. In this example, $\beta=1.027,  \Omega=0.2053, k_f=0.9998, k_g=0.8421, L=20.35, \Lambda=5.938$. \par
Last, we consider two limiting cases.  When $k_g\rightarrow 0$, $\beta$ attains the upper limit of (5.58) 
and other parameters behave like
$$\Omega \sim {A\over A^2-1},\ k_f \sim {\sqrt{A^4-1}\over A^2},
\ f \sim A\ {\rm dn}(\beta\xi, k_f), \ g \sim \cot \Omega\eta. \eqno(5.61)$$
The solution (5.59) reduces to
$$u\sim -{4A\over a}{\ \Omega\ {\rm dn}(\beta\xi, k_f)+\beta k_f^2\ {\rm sn}(\beta\xi, k_f)
{\rm cn}(\beta\xi, k_f){\rm sin} \Omega\eta\ {\rm cos} \Omega\eta
\over A^2{\rm dn}^2(\beta\xi, k_f){\rm sin}^2\Omega\eta + {\rm cos}^2\Omega\eta}, \eqno(5.62a)$$
$$x\sim y-{4\beta\over a}{1\over A^2{\rm dn}^2(\beta\xi, k_f){\rm sin}^2 \Omega\eta + {\rm cos}^2 \Omega\eta}
\Big[{1\over A}(1-A^2{\rm dn}^2(\beta\xi, k_f)){\rm sin}\ \Omega\eta\ {\rm cos}\ \Omega\eta $$
$$-k_f^2A^2{\rm sn}(\beta\xi, k_f){\rm cn}(\beta\xi, k_f){\rm dn}(\beta\xi, k_f){\rm sin}^2\Omega\eta\Big]
-{4\beta\over a}\left(E(\beta\xi, k_f) -{1\over A} \Omega\eta\right)+d. \eqno(5.62b)$$
In the case of $m_\xi=2$ and $m_\eta=1$, the relation (5.48) determines $A$ uniquely as $A=6.553$
so that $\beta=1.024, \Omega=0.1562, k_f=0.9997, L=20.11, \Lambda=5.669$. 
The solution (5.62) exhibits a profile similar to that depicted in Fig. 18. 
\par
\bigskip
    \begin{center}
\includegraphics[width=10cm]{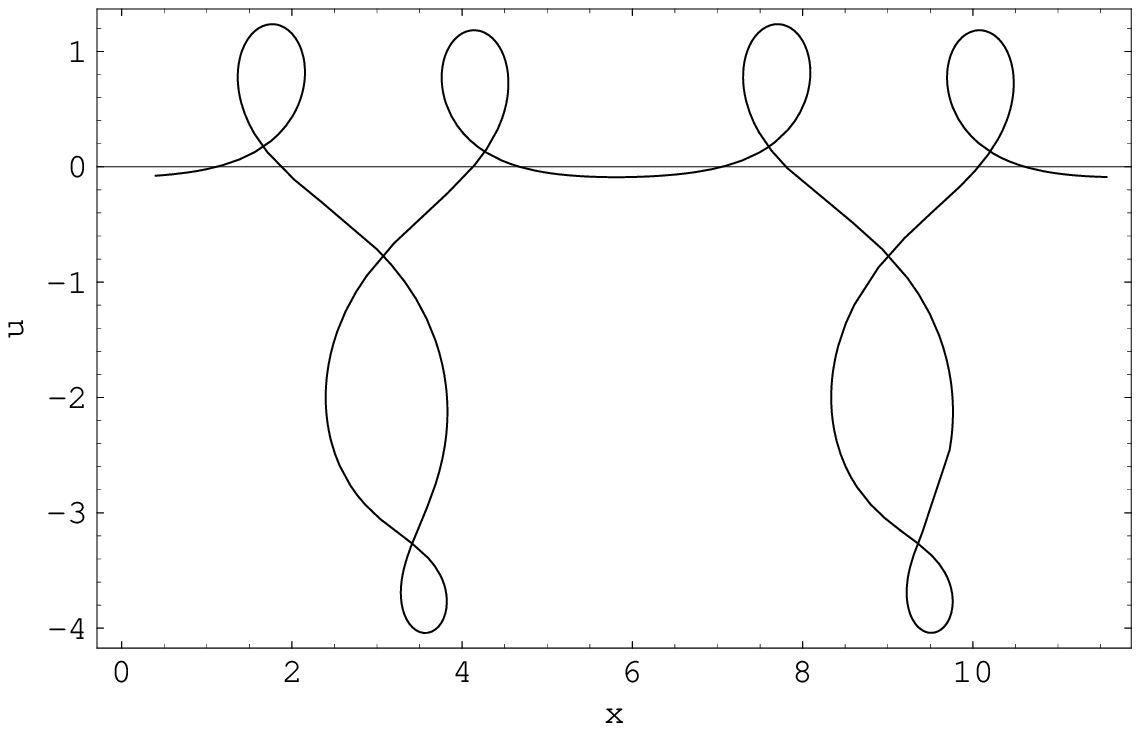}
\end{center}
\centerline{ {\bf Fig. 18}: A typical profile of the periodic solution for Example 3.} \par
\bigskip
The lower limit of $\beta$ in (5.58) is established when $k_f\rightarrow 1$ and $k_g\rightarrow 1$. Consequently, one has
$$\Omega\sim {1\over \sqrt{A^2-1}},\ f\sim A\ {\rm sech}\ \beta\xi, \ g\sim {\rm cosech}\ \Omega\eta. \eqno(5.63)$$
The solution then becomes
$$u \sim -{4\beta \over a}{\cosh \beta\xi\cosh \Omega\eta +A \sinh \beta\xi \sinh \Omega\eta
 \over  \cosh^2 \beta\xi +A^2 \sinh^2 \Omega\eta}, \eqno(5.64a)$$
 $$x \sim y -{2\beta\over a}{ \sinh 2\beta\xi -A \sinh 2\Omega\eta \over 
 \cosh^2 \beta\xi +A^2 \sinh^2 \Omega\eta} +d. \eqno(5.64b)$$
It represents the interaction between a loop soliton and an antiloop soliton.  See Fig. 3.
\par
\medskip
\leftline{\bf 5.3 Remarks}\par
\noindent 1. An elementary method for obtaining 1-phase solutions is available which reduces the SP
equation to a tractable ODE by assuming solution of traveling type [21]. \par
\noindent 2. The solutions (5.32), (5.42) and (5.56) have been derived in the context of a finite-length
sG system [22]. See also [23, 24] for analogous works. \par
\bigskip
\leftline{\bf 6 ALTERNATIVE METHOD OF SOLUTION}\par
\leftline{\bf 6.1 Bilinear transformation method}\par
The bilinear transformation method enables us to construct particular solutions of nonlinear
evolution equations [25-28]. Although this method has been employed to obtain soliton solutions
of the sG equation (see Sec. 4), it
is applicable to periodic solutions as well. Indeed, Nakamura developed a systematic procedure for constructing
periodic solutions of various types of soliton equations [29, 30]. Here, we shall use his method to
obtain periodic solutions of the sG equation. As  already demonstrated in Sec. 4 for constructing soliton solutions,
 the $\tau$-functions play an essential role in the bilinear formalism. In the periodic problem, we
 introduce the same dependent variable transformation as (4.1)
 $$\phi=2{\rm i}\, {\rm ln}\,{f^\prime\over f}. \eqno(6.1)$$
 Then, we can transform the sG equation (3.9) to the the following system of bilinear equations for the $\tau$-functions $f$ and $f^\prime$  
$$ff_{yt}-f_yf_t-{1\over 4}(f^2-{f^\prime}^2)=\lambda f^2,\eqno(6.2a)$$
$$f^\prime f^\prime_{yt}-f^\prime_y{f^\prime}_t-{1\over 4}({f^\prime}^2-f^2)=\lambda {f^\prime}^2,\eqno(6.2b)$$
where $\lambda$ is a complex parameter to be determined later and the variable $\tau$ has 
been replaced by the variable $t$ by virtue of (3.3a). 
The parametric representation of $u$ follows immediately from (3.8) and (6.1) 
$$u(y,t)=2{\rm i}\left(\ln {f^\prime\over f}\right)_t.\eqno(6.3)$$
We then use (6.1) and (6.2) to derive the relation
$$\cos \phi=1+4\lambda-2(\ln f^\prime f)_{yt}.\eqno(6.4)$$
 Introducing (6.4) into (3.14) and integrating with respect to $y$ yield the parametric representation of the
 coordinate $x$
$$x(y,t)=(1+4\lambda)y-2(\ln f^\prime f)_t+c. \eqno(6.5)$$
Comparing  (6.5) with (4.6), one sees that a new parameter $\lambda$ comes in the periodic solution which would disappear in the
long-wave (or soliton) limit. Thus, if we can solve the bilinear equations (6.2), then we can
obtain solutions of the SP equation through the parametric representation (6.3) and (6.5). 
It should be remarked that unlike the soliton solutions, the constant $c$ in (6.5) depends on $t$. This constant
can be determined by using (4.7). 
\par
\medskip
\leftline{\bf 6.2 Method of solution}\par
In accordance with Nakamura's procedure, we construct periodic solutions of the bilinear equations (6.2).
To this end, we first introduce the $N$-dimensional theta function
$$\theta({\bf z}|{\bf \tau})=\sum_{n_1,n_2,...,n_N=-\infty}^{\infty}
{\rm exp}\left(2\pi {\rm i}\sum_{j=1}^Nn_jz_j+\pi {\rm i}\sum_{j,k=1}^Nn_j\tau_{jk}n_k\right), \eqno(6.6)$$
where ${\bf z}=(z_1, z_2, ..., z_N)$ is an $N$-dimensional vector and ${\bf \tau}=(\tau_{jk})_{1\leq j,k\leq N}$ is
an $N\times N$ symmetric matrix. First, we seek solution of the bilinear equation (6.2a) in terms of the theta functions as
$$f=\theta\left({\bf z+{d\over 4}}\,\Big|{\bf \tau}\right), \eqno(6.7a)$$
$$ f^\prime=\theta\left({\bf z-{d\over 4}}\,\Big|{\bf \tau}\right), \eqno(6.7b)$$
where ${\bf d}=(1, 1, ..., 1)$ is an $N$-dimensional vector whose entries are all unity and $z_j\, (j=1, 2, ..., N)$ are phase variables
defined by
$$z_j=k_jy+\omega_jt+z_{j0}, \ (j=1, 2, ..., N). \eqno(6.7c)$$
Here, $k_j, \omega_j$ and $z_{j0}$ are complex parameters. Substituting (6.7) into (6.2a), we find that the
bilinear equation can be transformed to the form
$$\sum_{m_1,m_2,...,m_N=-\infty}^{\infty}F(m_1, m_2, ..., m_N){\rm exp}\left(2\pi {\rm i}\sum_{j=1}^Nm_jz_j\right)=0, \eqno(6.8a)$$
where
$$F(m_1, m_2, ..., m_N)$$
$$=\sum_{n_1,n_2,...,n_N=-\infty}^{\infty}
\Biggl[-2\pi^2\left\{\sum_{j=1}^N(2n_j-m_j)k_j\right\}\left\{\sum_{l=1}^N(2n_l-m_l)\omega_l\right\}$$
$$-{1\over 4}\left\{1+4\lambda-(-1)^{\sum_{j=1}^Nm_j}\right\}\Biggr]\times$$
$$\times {\rm exp}\left[\pi {\rm i}\left\{\sum_{j,k=1}^Nn_j\tau_{jk}n_k+\sum_{j,k=1}^N(m_j-n_j)\tau_{jk}(m_k-n_k)\right\}
+{\pi\over 2}{\rm i}\sum_{j=1}^Nm_j\right]. \eqno(6.8b)$$
By shifting the $s$th summation index $n_s$ as $n_s+1$ in (6.8b), we see that
$$F(m_1, m_2, ..., m_N)=-F(m_1, ..., m_{s-1}, m_s-2, m_{s+1}, ..., m_N)\times$$
$$\times {\rm exp}\left[2\pi {\rm i}\left(\sum_{l=1}^N\tau_{sl}m_l-\tau_{ss}\right)\right]. \eqno(6.9)$$
Thus, if the relations
$$F(m_1, m_2, ..., m_N)=0, \eqno(6.10)$$
hold for all possible combinations of $m_1=0, 1, m_2=0, 1, ..., m_N=0, 1$, then all $F's$ become zero for
arbitrary integer values of $m_1, m_2, ..., m_N$, implying that  Eq. (6.8) holds identically.
Consequently, the $\tau$-functions (6.7) satisfy the bilinear equation (6.2a).\par
A similar analysis shows that the bilinear equation (6.2b)  reduces to Eq. (6.8a) where the function $F$ has the same form
as (6.8b) except that the factor ${\pi\over 2}{\rm i}\sum_{j=1}^Nm_j$ in the exponential function is replaced simply by
$-{\pi\over 2}{\rm i}\sum_{j=1}^Nm_j$. It turns out that the relations (6.10) assure that the $\tau$-functions (6.7)
satisfy the bilinear equation (6.2b) as well. Thus, if we can determine parameters such that relations (6.10)
are satisfied, then we obtain
periodic solutions of the sG equation. 
We can regard (6.10) as a system of $2^N$ nonlinear  equations for the unknown parameters $\omega_j (j=1, 2, ..., N),
\tau_{jk} (1\leq j<k\leq N)$ and $\lambda$ with given values of $k_j$ and $\tau_{jj} (j =1, 2, ., N)$.
The total number of unknowns is $N(N-1)/2+N+1=N(N+1)/2+1$. For $N=1, 2$, the total number of equations
is equal to the total number of unknown parameters. Hence, we  have 1- and 2-phase solutions
in terms of the theta functions. For $N\geq 3$, on the other hand, the total number of equations always exceeds that
of unknowns. In this case, Eqs. (6.10) become an overdetermined system and we need a separate consideration
as for the existence of the solution. \par
\medskip 
\leftline{\bf 6.3 1-phase solutions}\par
Here, we derive a 1-phase solution of the sG equation by means of the method described above. For $N=1$, the
relations (6.10) become
$$F(m)=\sum_{n=-\infty}^\infty\left[-2\pi^2(2n-m)^2k\omega-{1\over 4}\left\{1+4\lambda-(-1)^m\right\}\right]\times$$
$$\times {\rm exp}\left[\pi {\rm i}\{(n-m)^2+n^2\}\tau+{\pi\over 2}{\rm i}m\right]=0, \ (m=0, 1), \eqno(6.11)$$
where we have put $k=k_1, \omega=\omega_1, \tau=\tau_{11}, m=m_1, n=n_1$ for simplicity.
Explicitly, these read 
$$\sum_{n=-\infty}^\infty(8\pi^2n^2k\omega+\lambda){\rm e}^{2\pi {\rm i}n^2\tau}=0, \eqno(6.12a)$$
$$\sum_{n=-\infty}^\infty\{2\pi^2(2n-1)^2k\omega+{1\over 2}(1+2\lambda)\}{\rm e}^{\pi {\rm i}\{(n-1)^2+n^2\}\tau}=0. \eqno(6.12b)$$
We can rewrite (6.12) in terms of the following 1-dimensional theta functions 
$$\theta_1(z|\tau)=-{\rm i}\sum_{n=-\infty}^\infty (-1)^n{\rm exp}\left[\pi {\rm i}(2n+1)z+\pi {\rm i}\left(n+{1\over 2}\right)^2\tau\right], \eqno(6.13a)$$
$$\theta_2(z|\tau)=\sum_{n=-\infty}^\infty {\rm exp}\left[\pi {\rm i}(2n+1)z+\pi {\rm i}\left(n+{1\over 2}\right)^2\tau\right], \eqno(6.13b)$$
$$\theta_3(z|\tau)=\sum_{n=-\infty}^\infty {\rm exp}\left[2\pi {\rm i}nz+\pi {\rm i}n^2\tau\right], \eqno(6.13c)$$
$$\theta_4(z|\tau)=\sum_{n=-\infty}^\infty (-1)^n{\rm exp}\left[2\pi {\rm i}nz+\pi {\rm i}n^2\tau\right]. \eqno(6.13d)$$
Before proceeding, it is convenient to introduce a new parameter $q$ by
$$q={\rm e}^{\pi {\rm i}\tau}, \eqno(6.14)$$
and write the above four theta functions as
$$\theta_j(z|\tau)=\theta_j(z,q), \ (j=1, 2, 3, 4). \eqno(6.15)$$
Thus, the relation $\theta_j(z|n\tau)=\theta_j(z,q^n)$ holds for
any integer $n$. This notation will be used in the following.  \par
Now, using (6.15), Eqs. (6.12) can be recast into the  following system of linear algebraic equation for the
two unknowns $k\omega$ and $\lambda$
$$2\theta_3^{\prime\prime}(0,q^2)k\omega-\theta_3(0,q^2)\lambda=0, \eqno(6.16a)$$
$$2\theta_2^{\prime\prime}(0,q^2)k\omega-{1\over 2}\theta_2(0,q^2)(1+2\lambda)=0, \eqno(6.16b)$$
where $\theta_j^{\prime\prime}(0,q^2)=d^2\theta_j(z,q^2)/dz^2|_{z=0},\ (j=2, 3)$. Solving this system, we obtain
$$k\omega=-{1\over 4}{\theta_2(0,q^2)\theta_3(0,q^2)\over \theta_2(0,q^2)\theta_3^{\prime\prime}(0,q^2)
-\theta_2^{\prime\prime}(0,q^2)\theta_3(0,q^2)}, \eqno(6.17a)$$
$$\lambda=-{1\over 2}{\theta_2(0,q^2)\theta_3^{\prime\prime}(0,q^2)\over \theta_2(0,q^2)\theta_3^{\prime\prime}(0,q^2)
-\theta_2^{\prime\prime}(0,q^2)\theta_3(0,q^2)}. \eqno(6.17b)$$
If we use the identity
$${\theta_3^{\prime\prime}(0,q^2)\over \theta_3(0,q^2)}-{\theta_2^{\prime\prime}(0,q^2)\over \theta_2(0,q^2)}
=\pi^2\theta_4^4(0,q^2), \eqno(6.18)$$
we can recast (6.17) to the compact expressions
$$k\omega=-{1\over 4\pi^2\theta_4^4(0,q^2)},\eqno(6.19a)$$
$$\lambda=-{\theta_3^{\prime\prime}(0,q^2)\over 2\pi^2\theta_3(0,q^2)\theta_4^4(0,q^2)}. \eqno(6.19b)$$
\par
Now, the $\tau$-functions $f$ and $f^\prime$ can be expressed in the form
$$f = \theta\left(z+{1\over 4}\,\Big|\tau \right) 
   = \sum_{n=-\infty}^\infty{\rm exp}\left[2\pi {\rm i}n\left(z+{1\over 4}\right)+\pi {\rm i}n^2\tau\right], 
  \eqno(6.20a)$$
  $$f^\prime = \theta\left(z-{1\over 4}\,\Big|\tau \right) 
   = \sum_{n=-\infty}^\infty{\rm exp}\left[2\pi {\rm i}n\left(z-{1\over 4}\right)+\pi {\rm i}n^2\tau\right], 
  \eqno(6.20b)$$
  with
  $$z=ky+\omega t+z_0. \eqno(6.20c)$$
  In order to obtain a real periodic solution, we introduce the new real quantities with tilde by
  $$k=-{{\rm i}\over 2\pi}\tilde k,\ \omega=-{{\rm i}\over 2\pi}\tilde\omega,\ z_0=-{{\rm i}\over 2\pi}\tilde z_0, \eqno(6.21a)$$
  $$  z=-{{\rm i}\over 2\pi}\tilde z=-{{\rm i}\over 2\pi}(\tilde ky+\tilde\omega t+\tilde z_0), \eqno(6.21b)$$
  and put
  $$\tau={\rm i}b, \ (b>0), \eqno(6.21c)$$
  to assure the convergence of the series (6.20). Then, $f$ and $f^\prime$ are rewritten as
  $$f 
   = \sum_{n=-\infty}^\infty{\rm exp}\left[n\left(\tilde z+{\pi\over 2}{\rm i}\right)-\pi n^2b \right], 
  \eqno(6.22a)$$
  $$f^\prime 
   = \sum_{n=-\infty}^\infty{\rm exp}\left[n\left(\tilde z-{\pi\over 2}{\rm i}\right)-\pi n^2b \right],
  \eqno(6.22b)$$
  with
  $$\tilde z=\tilde ky+\tilde\omega t+\tilde z_0. \eqno(6.22c)$$
  In terms of the new parameters $\tilde\omega$ and $\tilde k$, the dispersion relation (6.19a) becomes
  $$\tilde\omega={1\over \theta_4^4(0,q^2)\tilde k},\ (q={\rm e}^{-\pi b}). \eqno(6.23)$$
  Obviously, $f^\prime=f^*$ and $\lambda$ is  real. Hence, the parametric solution given by (6.3) and (6.5) yields a
  real 1-phase periodic solution of the SP equation. For computing $u$ from (6.3), we rewrite $f$ in terms of the 
  theta functions $\theta_1$ and $\theta_4$ as
  $$f=\theta_4\left({\tilde z\over \pi {\rm i}},q^4\right)
  +{\rm i}\left\{{\rm i}\theta_1\left({\tilde z\over \pi {\rm i}},q^4\right)\right\}. \eqno(6.24a)$$
  Note that $\theta_4(\tilde z/\pi {\rm i},q^4)$ and ${\rm i}\theta_1(\tilde z/\pi {\rm i},q^4)$ are real functions of $\tilde z$.
  The $\tau$-function $f^\prime$ is given by the complex conjugate of $f$. It reads 
  $$f^\prime=\theta_4\left({\tilde z\over \pi {\rm i}},q^4\right)
  -{\rm i}\left\{{\rm i}\theta_1\left({\tilde z\over \pi {\rm i}},q^4\right)\right\}. \eqno(6.24b)$$
  It follows from (6.24a) and the definition of the sn function in terms of the theta functions that
  $${{\rm Im}\, f\over {\rm Re}\, f}={{\rm i}\theta_1\left({\tilde z\over \pi {\rm i}},q^4\right)
  \over \theta_4\left({\tilde z\over \pi {\rm i}},q^4\right)}
  ={\rm i}\,{\theta_2(0,q^4)\over \theta_3(0,q^4)}\, {\rm sn}(v,\kappa), \eqno(6.25a)$$
  where
  $$v=-{\rm i}\,\theta_3^2(0,q^4)\tilde z,\ \kappa={\theta_2^2(0,q^4)\over \theta_3^2(0,q^4)}. \eqno(6.25b)$$
  Furthermore, using the formula
  $${\rm sn}({\rm i}v,\kappa)={\rm i}\,{{\rm sn}(v,\kappa^\prime)\over {\rm cn}(v,\kappa^\prime)},\ \kappa^\prime=\sqrt{1-\kappa^2}
  ={\theta_4^2(0,q^4)\over \theta_3^2(0,q^4)}, \eqno(6.26)$$
  (6.25) becomes
  $${{\rm Im}\, f\over {\rm Re}\, f}=
  {\theta_2(0,q^4)\over \theta_3(0,q^4)}{{\rm sn}(\theta_3^2(0,q^4)\tilde z, \kappa^\prime)
  \over {\rm cn}(\theta_3^2(0,q^4)\tilde z, \kappa^\prime)}. \eqno(6.27)$$
  The relation $f^\prime=f^*$ makes it possible to write (6.3) as $u=4[\tan^{-1}({\rm Im}\,f/{\rm Re}\, f)]_t$.
  Substitution of (6.27) into this expression yields $u$ given by (5.51)
   with the identification
  among the parameters
  $$A={\theta_2(0,q^4)\over \theta_3(0,q^4)},\ a=\theta_4^2(0,q^2)\tilde k, \ \beta={\theta_3^2(0,q^4)\over \theta_4^2(0,q^2)},
   \ \kappa^\prime=\sqrt{1-A^4}. \eqno(6.28)$$
  \par
    To derive the expression of $x$ from (6.5), on the other hand, one needs the time dependence of $c$ which can
    be determined from (6.24) and (4.7). After some calculations using
    the identities of the theta functions, we find
    $$c(t)=-{4\tilde\omega^2\over \pi^2}\left[{\theta_4^{\prime\prime}(0,q^4)\over\theta_4(0,q^4)}
    +\pi^2\theta_2^2(0,q^4)\theta_3^2(0,q^4)\right]t+d, \eqno(6.29)$$
    where $d$ is an arbitrary real constant. It can be demonstrated  by substituting (6.24) and (6.29) into (6.5)
    that the expression (6.5) coincides with (5.52). 
    Indeed, a straightforward calculation using (6.5), (6.20), (6.28) and some formulas for the theta
    functions and Jacobi elliptic functions leads to the expression of $x$. We write it in the form
    $$x=y+{4\over a}{(1+A^2)\, {\rm sn}(\beta\xi, k_f){\rm cn}(\beta\xi, k_f){\rm dn}(\beta\xi, k_f)
\over {\rm dn}^2(\beta\xi, k_f)+A^2}$$
$$+{4\beta\over a}\left(-E(\beta\xi, k_f)
+{\theta_4^{\prime\prime}(0,q^4)\over \pi^2\theta_3^2(0,q^4)\theta_4(0,q^4)}\,\tilde z\right)+4\lambda y+c(t). \eqno(6.30)$$
Note from (3.10a), (6.22c), (6.23) and (6.28) that $\tilde z=\beta\xi/\theta_3(0,q^4)$ and from (3.10)
that the last two terms on the right-hand side of (6.30) can be expressed in terms of $\xi$ and $\eta$. With these facts in mind,
we then use the formulas
$${\theta_3^{\prime\prime}(0,q^2)\over \theta_3^2(0,q^2)}
={2\theta_4(0,q^4)\theta_4^{\prime\prime}(0,q^4)\over \theta_3(0,q^2)\theta_4(0,q^2)}-{\pi^2\over 2}\theta_2^4(0,q^2), \eqno(6.31a)$$
$$\theta_2(0,q^4)\theta_3(0,q^4)={1\over 2}\theta_2^2(0,q^2), \eqno(6.31b)$$
and see that (6.30) coincides perfectly with the expression of $x$ given by (5.52).
    \par
    Last, we consider the soliton limit. To this end, we first shift the phase constant $\tilde z_0$
    as $\tilde z_0\rightarrow \tilde z_0+\pi b$ and take the limit $b\rightarrow \infty$
    (or $q\rightarrow 0$). In this limit, the theta functions $\theta_3$ and $\theta_4$ have
    the power series expansions
    $$\theta_3(0,q)=1+2q+O(q^4), \eqno(6.32a)$$
    $$\theta_4(0,q)=1-2q+O(q^4). \eqno(6.32b)$$
Then, the asymptotic form of $\lambda$ from (6.17b) and that  of $\tilde \omega$ from (6.23) become
$$\lambda\sim 0,\ \tilde\omega\sim {1\over \tilde k}. \eqno(6.33)$$
The $\tau$-function from (6.22a) behaves like
$$f\sim 1+{\rm i}{\rm e}^{\tilde z},\ \tilde z=\tilde ky+{1\over\tilde k}t+\tilde z_0. \eqno(6.34)$$
This coincides with the $\tau$-function (4.9) for the 1-loop soliton solution. \par
Another type of real 1-phase solutions can be constructed by a similar procedure to that described above.
We list two of them for reference. If we replace $z$ by $z+{1\over 4}$ and put $\tau={{\rm 1}\over 2}+{\rm i}b$ in (6.20), 
the $\tau$-functions $f$ and $f^\prime$ turn out to be
$$f     = \sum_{n=-\infty}^\infty{\rm exp}\left[2\pi {\rm i}nz-{\pi {\rm i}\over 2}n^2 -\pi {\rm i}n^2\tau\right], 
  \eqno(6.35a)$$
  $$f^\prime    =\sum_{n=-\infty}^\infty{\rm exp}\left[2\pi {\rm i}nz+{\pi {\rm i}\over 2}n^2 -\pi {\rm i}n^2\tau\right] , 
  \eqno(6.35b)$$
where we have used the formula $e^{\pi {\rm i}n(n+1)}=1$.  
In view of the formulas $\theta_3(z|\tau+1)=\theta_4(z,\tau), \theta_4(z|\tau+1)=\theta_3(z,\tau)$, the relations (6.19) then become
$$k\omega=-{1\over 4\pi^2\theta_3^4(0,q^2)},\eqno(6.36a)$$
$$\lambda=-{\theta_4^{\prime\prime}(0,q^2)\over 2\pi^2\theta_4(0,q^2)\theta_3^4(0,q^2)}. \eqno(6.36b)$$
Using (6.35a), we find
$${{\rm Im}\, f\over {\rm Re}\, f}=-A\,{{\rm cn}(w,\kappa)\over {\rm dn}(w,\kappa)}, \eqno(6.37a)$$
where
$$w=2\pi^2\theta_3^2(0,q^4)(ky+\omega t+z_0), \ A={\theta_2(0,q^4)\over \theta_3(0,q^4)},\ \kappa=A^2,
\ q={\rm e}^{-\pi b}. \eqno(6.37b)$$
\par
By replacing $z$ by ${\rm i}z$ in (6.35), we obatin 
$${{\rm Im}\, f\over {\rm Re}\, f}=-{A\over {\rm dn}(w,\kappa^\prime)}, \eqno(6.38)$$
with $\kappa^\prime=\sqrt{1-\kappa^2}$, where the expressions (6.36) remain the same forms. 
The expressions (6.37) and (6.38) give rise to the 1-phase solutions of the sG equation. \par 
\medskip
\leftline{\bf 6.4 2-phase solutions}\par
In order to construct 2-phase solutions of the sG equation, we first solve the system of equations
(6.10). Given  values of $k_1, k_2, \tau_{11}$ and $\tau_{22}$, this system can be solved in principle
for the unknown parameters $\omega_1, \omega_2, \tau_{12} $ and $\lambda.$
Because of the transcendental nature of the system of equations, however, it is very difficult to obtain
analytical solution. Under a few special situations in which the 2-dimensional theta function is
expressed by a finite sum of 1-dimensional theta functions, the system can be solved algebraically.
Indeed, there exist several examples to realize the situation mentioned above [31]. Among them,
we consider the particularly important case
$$\tau_{11}=\tau_{22}. \eqno(6.39)$$
It turns out that the 2-dimensional theta function (6.6) has the representation [31]
$$\theta({\bf z}|\tau)={1\over 2}\left\{\theta_3\left(z_++{1\over 2}\Big|\tau_+\right)\theta_3\left(z_-+{1\over 2}\Big|\tau_-\right)
+\theta_3\left(z_+|\tau_+\right)\theta_3\left(z_-|\tau_-\right)\right\}, \eqno(6.40a)$$
where
$$z_\pm={1\over 2}(z_1\pm z_2), \eqno(6.40b)$$
$$\tau_\pm={1\over 2}(\tau_{11}\pm\tau_{12}), \ ({\rm Im}\,\tau_\pm>0). \eqno(6.40c)$$
It follows from (6.7) and (6.40) that
$$f=\theta_3\left(z_+-{1\over 4}\Big|\tau_+\right)\theta_3\left(z_-+{1\over 2}\Big|\tau_-\right)
+\theta_3\left(z_++{1\over 4}\Big|\tau_+\right)\theta_3\left(z_-\big|\tau_-\right),\eqno(6.41a)$$
$$f^\prime=\theta_3\left(z_++{1\over 4}\Big|\tau_+\right)\theta_3\left(z_-+{1\over 2}\Big|\tau_-\right)
+\theta_3\left(z_+-{1\over 4}\Big|\tau_+\right)\theta_3\left(z_-\big|\tau_-\right).\eqno(6.41b)$$
Invoking the definition of the dn function, the solution $\phi$ of the sG equation can be written in the form
$${\rm tan}\,{\phi\over 4}={1\over {\rm i}}{f-f^\prime\over f+f^\prime}$$
$$={\sqrt{k_+^\prime}-{\rm dn}(u_++\delta_+,k_+)\over \sqrt{k_+^\prime}+{\rm dn}(u_++\delta_+,k_+)}
{\sqrt{k_-^\prime}-{\rm dn}(u_-,k_-)\over \sqrt{k_-^\prime}+{\rm dn}(u_-,k_-)}, \eqno(6.42a)$$
where
$$u_\pm=\pi \theta_3^2(0|\tau_\pm)z_\pm, \eqno(6.42b)$$
$$k_\pm=\sqrt{1-{k_\pm^\prime}^2}, \ k_\pm^\prime={\theta_4^2(0|\tau_\pm)\over \theta_3^2(0|\tau_\pm)}, \eqno(6.42c)$$
$$\delta_+={\pi\over 4}\theta_3^2(0|\tau_+). \eqno(6.42d)$$
Thus, the above solution has a form similar to (5.23) in which the variables $u_+$ and $u_-$ are separated completely. \par
The next step is to solve the system of equations (6.10) with $N=2$. We observe under the condition (6.39) that
$$n_1^2\tau_{11}+2n_1n_2\tau_{12}+n_2^2\tau_{22}=(n_1+n_2)^2\tau_++(n_1-n_2)^2\tau_-. \eqno(6.43)$$
Using (6.43), Eqs. (6.10) can be solved analytically. The calculation involved is straightforward but 
somewhat tedious. Hence,
we outline only the main steps. It follows from the equations $F(1, 0)=0$ and $F(0, 1)=0$ that
$$k_1\omega_1=k_2\omega_2. \eqno(6.44)$$
Substituting $\omega_2$ from (6.44) into the equations $F(0, 0)=0$ and $F(1, 1)=0$, one obtains a homogeneous system of linear
equations for $\omega_1$ and $\lambda$. The solvability of this sytem yields the relation
$$[\theta_2^{\prime\prime}(0,q_+^8)\theta_3(0,q_+^8)-\theta_2(0,q_+^8)\theta_3^{\prime\prime}(0,q_+^8)]
[\theta_2^2(0,q_-^8)-\theta_3^2(0,q_-^8)]$$
$$=\alpha [\theta_2^{\prime\prime}(0,q_-^8)\theta_3(0,q_-^8)-\theta_2(0,q_-^8)\theta_3^{\prime\prime}(0,q_-^8)]
[\theta_2^2(0,q_+^8)-\theta_3^2(0,q_+^8)], \eqno(6.45a)$$
where
$$q_\pm={\rm e}^{\pi {\rm i}\tau_\pm}={\rm e}^{\pi {\rm i}(\tau_{11}\pm\tau_{12})/2}, \eqno(6.45b)$$
$$\alpha=\left({k_1-k_2\over k_1+k_2}\right)^2. \eqno(6.45c)$$
Through a sequence of transformations using various formulas of the theta functions, one can simplify (6.45) to the form
$$\theta_2^4(0,q_+^2)=\alpha\theta_2^4(0,q_-^2). \eqno(6.46)$$
This is a transcendental equation which determines $q_+$ for given $q_-$ and $\alpha$. 
Once $q_+$ is obtained, the parameter $\tau_{12}$
is found from the relation
$${\rm e}^{\pi {\rm i}\tau_{12}}={q_+\over q_-}, \eqno(6.47)$$
which follows from (6.45b).
The parameters $\omega_1$ and $\lambda$ are then determined from the equations $F(0, 0)=0$ and $F(1, 0)=0$ 
as well as  the relation (6.44), the explicit forms of which are not written down here. Thus, we have completed
the construction of a 2-phase solution of the bilinear equation (6.2a). The corresponding solution of the SP
equation can be obtained from (6.1) and (6.5). As in the case of 1-phase solutions discussed in Sec. 6.3,
various 2-phase solutions of the SP equation arise by specifying the parameters $k_1, k_2,$ and $\tau_{11}(=\tau_{22})$.
The  explicit solutions are not presented here and will reported elsewhere.\par
\medskip
\leftline{\bf 6.5 Remarks}\par
\noindent 1. The starting point of our discussion
is the condition (6.39)  which makes it possible to perform all the calculations algebraically. The resulting solution
of $\phi$ has a separable form with respect to the independent variables $u_+$ and $u_-$ (see (6.42)).  This expression
should be compared with the separable solution  introduced in (5.23).  Thus, we can expect that the method developed here
would reproduce all solutions constructed by a  different method described in Sec. 5.2. \par
\noindent 2. Unless the condition (6.39) is
imposed, we would be able to obtain a broader class of 2-phase solutions. The structure of solutions is worth
studying in a future work. \par
\noindent 3. The general $N$-phase solution of the sG equation has been constructed by means
of the method of algebraic geometry [18]. It has a form given by (6.1) in terms of the $N$-dimensional theta functions.
However, whether the corresponding $\tau$-functions satisfy a system of  bilinear equations (6.2) or not
is an open problem to be resolved in a different context.
On the other hand, our method first looks for the solution of (6.2) so that the expression of $x$ follows immediately
from (6.5). As already mentioned, however, the construction of the $N$-phase solutions with $N \geq 3$ is a difficult task 
in the context of the bilinear formalism. Nevertheless, it is undoubtedly a challenging problem. \par
\bigskip
\leftline{\bf 7 CONCLUSION}\par
We have presented soliton and periodic solutions of the SP equation by means of a new method of
solution. The most difficult technical problem in constructing solutions was how to integrate
the PDE which governs the inverse mapping to the original coordinate system (see (3.13) and (3.14)).
In the case of soliton solutions, the explicit form of the coordinate $x$ was obtained in terms of
the $\tau$-functions $f$ and $f^\prime$ as shown by (4.5). In the case of periodic solutions, on
the other hand, the special ansatz leads to the explicit form of $x$ in terms of  Jacobi's
elliptic functions. Specifically, for 1-phase solutions, assuming the dependence of single
variable, the sG equation becomes a tractable ODE (5.1) so that a number of special solutions are
available. For 2-phase solutions, under the ansatz (5.23), we were able to reduce the representation
of $x$  to single integrals which are easily integrated (see (5.31)).  It should be remarked, however,
that the resulting solutions of the SP equation are special class of real 2-phase solutions.
An alternative method employing the bilinear transformation method described in Sec. 6 enables us to
construct a broader class of solutions than the solutions obtained in Sec. 5. At present, the most interesting
issue will be the construction of the general $N$-phase solution which is reduced to the
$N$-soliton solution (4.5) and (4.6) in the long-wave limit.\par
\bigskip
\leftline{\bf REFERENCES}\par
\begin{enumerate}[{[1]}]
\item T. Sch\"afer and C.E. Wayne, {\it Propagation of ultra-short optical pulses in cubic
    nonlinear media}, Physica {\bf D 196} (2004) 90-105.
\item J.E. Rothenberg, {\it Space-time focusing: breakdown of the slowly varying envelope approximation
    in the self-focusing of femtosecond pulses}, Opt. Lett. {\bf 17} (1992) 1340-1342. 
\item J.K. Ranka and A.L. Gaeta, {\it Breakdown of the slowly varying envelope approximation in the
    self-focusing of ultrashort pulses}, Opt. Lett. {\bf 23} (1988) 534-536.
\item Y. Chung, C.K.R. Jones, T. Sch\"afer and C.E. Wayne, {\it Ultra-short pulses in linear and
    nonlinear media}, Nonlinearity, {\bf 18} (2005) 1351-1374.
\item M.L. Rabelo. {\it On equations which describe pseudospherical surfaces},
    Stud. Appl. Math. {\bf 81} (1989) 221-248.
\item D. Alterman and J. Rauch, {\it Diffractive short pulse asymptotics for nonlinear wave equations},
    Phys. Lett. {\bf A 264} (2000) 390-395.
\item R. Beals, M. Rabelo and K. Tenenblat, {\it B\"acklund transformations and inverse scattering solutions
   for some pseudospherical surface equations}, Stud. Appl. Math. {\bf 81} (1989) 121-151.
\item A. Sakovich and S. Sakovich, {\it The short pulse equation is integrable}, 
   J. Phys. Soc. Jpn. {\bf 74} (2005) 239-241.
\item J.C. Brunelli, {\it The short pulse hierarchy}, J. Math. Phys. {\bf 46} (2005) 123507(1-9).
\item J.C. Burunelli, {\it The bi-Hamiltonian structure of the short pulse equation},
   Phys. Lett. {\bf A 353} (2006) 475-478.
\item Y. Matsuno, {\it Cusp and loop soliton solutions of short-wave models for the Camassa-Holm
   and Degasperis-Procesi equations}, 
   Phys. Lett. {\bf A 359} (2006) 451-457.
\item Y. Matsuno, {\it Multisoliton and multibreather solutions of the short pulse model equation},
   J. Phys. Soc. Jpn. {\bf 76} (2007) 084003(1-6).
\item R. Hitota, {\it Exact solution of the sine-Gordon equation for multiple collisions of solitons},
   J. Phys. Soc. Jpn. {\bf 33} (1972) 1459-1463.
\item A. Sakovich and S. Sakovich, {\it Solitary wave solutions of the short pulse equation},
    J. Phys. A: Math. Gen. {\bf 39} (2006) L361-L367.
\item V.K. Kuetche, T.B. Bouetou and T.C. Kofane, {\it On two-loop soliton solution of the
    Sch\"afer-Wayne short-pulse equation using Hirota's method and Hodnett-Moloney approach},
    J. Phys. Soc. Jpn. {\bf 76} (2007) 024004(1-7).
\item Y. Matsuno, {\it Periodic solutions of the short pulse model equation}, 
    J. Math. Phys. {\ 49} (2008) 073508 (1-18).
\item D.F. Lawden, {\it Elliptic Functions and Applications} (Springer, New York, 1989).
\item E.D. Belokolos, A.I. Bobenko, V.Z. Enol'skii, A.R. Its and V.B. Matveev,
    {\it Algebro-Geometric Approach to Nonlinear Integrable Equations} (Springer, New York, 1994).
\item G.L. Lamb, Jr., {\it Analytical descriptions of ultrashort optical pulse propagation in a
    resonant medium}, Rev. Mod. Phys. {\bf 43} (1971) 99-124.
\item G.L. Lamb, Jr., {\it Elements of Soliton Theory} (Wiley, New York, 1980).
\item E.J. Parkes, {\it Some periodic and solitary travelling-wave solutions of the short-pulse equation},
    Chaos, Solitons and Fractals {\bf 38} (2008) 154-159.
\item G. Costabile, R.D. Parmentier, B. Savo, D.W. McLaughlin and A.C. Scott,
{\it Exact solutions of the sine-Gordon equation describing oscillations in a long (but finite) 
Josephson junction}, Appl. Phys. Lett. {\bf 32} (1978) 587-589.
\item R.M. DeLeonardis and S.E. Trullinger, {\it Theory of boundary effects on sine-Gordon solitons},
J. Appl. Phys. {\bf 51} (1980) 1211-1226.
\item R.M. DeLeonardis, S.E. Trullinger and R.F. Wallis, {\it Classical excitation energies for a
finite-length sine-Gordon system},
J. Appl. Phys. {\bf 53} (1982) 699-702.
\item R. Hirota, {\it Exact solution of the Korteweg-de Vries equation for multiple collision of solitons},
Phys. Rev. Lett. {\bf 27} (1971) 1192-1194.
\item R. Hirota, {\it Direct Method in Soliton Theory}, in {\it Solitons}, R.K. Bullough 
and P.J. Caudrey eds. (Springer, New York, 1980) 157-176.
\item Y. Matsuno, {\it Bilinear Transformation Method} (Academic Press, New York, 1984).
\item R. Hirota, {\it Direct Method in Soliton Theory} (Cambridge University Press, Cambridge, 2004).
\item A. Nakamura, {\it A direct method of calculating periodic wave solutions to nonlinear 
evolution equations. I. Exact two-periodic wave solution},
 J. Phys. Soc. Jpn. {\bf 47} (1979) 1701-1705.
 \item A. Nakamura, {\it A direct method of calculating periodic wave solutions to nonlinear 
evolution equations. II. Exact one- and two-periodic wave solution of the coupled bilinear equations},
 J. Phys. Soc. Jpn. {\bf 48} (1980) 1365-1370.
\item J. Zagrodzi\'nski, {\it Dispersion equations and a comparison of different quasi-periodic solutions
of the sine-Gordon equation}, J. Phys. A: Math. Gen. {\bf 15} (1982) 3109-3118.

\end{enumerate} 
\end{document}